\documentclass[a4paper,11pt]{article}
\usepackage{pos}

\title{Cataclysmic Variables from {\it XMM-Newton} and eROSITA}

\author*[a]{Daniela Muñoz-Giraldo}
\author[a]{Beate Stelzer}
\author[b]{Axel Schwope}

\affiliation[a]{Institut f\"ur Astronomie und Astrophysik, Eberhard Karls Universit\"at T\"ubingen,\\
  Sand 1, 72076 T\"ubingen, Germany}
\affiliation[b]{Leibniz-Institut f\"ur Astrophysik Potsdam (AIP),\\
An der Sternwarte 16, 14482 Potsdam, Germany}

\emailAdd{munoz-giraldo@astro.uni-tuebingen.de}

\abstract{Cataclysmic variables (CVs) are interacting binaries in which a white dwarf accretes material from a late-type companion, with the most evolved systems, commonly referred to as period-bouncers, hosting degenerate donors. We present results from two major X-ray facilities, {\it XMM-Newton} and eROSITA, that provide complementary views of the CV population. We first summarize the capabilities of {\it XMM-Newton} for timing and spectroscopic studies, highlighting its ability to characterize accretion geometries, magnetic field strengths, and short-term variability across both magnetic and non-magnetic CVs. These data enable detailed studies of individual systems, including magnetic period-bouncers, with soft X-ray spectra and faint luminosities that offer insights into the late stages of CV evolution.

We then present an overview of eROSITA as the first X-ray all-sky survey in the soft band since ROSAT, emphasizing its impact on population studies. The eRASS:4 catalog dramatically expands the known X-ray sky and will enable the construction of volume-limited samples of CVs, complete up to 500pc for standard CVs and up to 250 pc for period-bouncers. Cross-matching eROSITA detections with optical catalogs has already produced the largest sample of X-ray–selected CV candidates to date, that will allow us to refine X-ray diagnostic diagrams, reassess the demographics of magnetic and non-magnetic systems, and identify dozens of new highly evolved and faint CVs.

Finally, we describe ongoing searches for previously unknown systems, including CVs selected through probabilistic cross-matching with Gaia, and period-bouncers identified through a multi-wavelength scorecard combining optical, infrared, ultraviolet, and X-ray properties. These efforts have already led to the discovery of new period-bouncers and a promising detached period-bouncer candidate, that could validate alternative models to describe the evolution of CVs. Together, {\it XMM-Newton} and eROSITA provide an unprecedented window into the structure, evolution, and demographics of CVs, with eROSITA having layed the groundwork for future studies of compact binary evolution, the Galactic Ridge X-ray emission, and related phenomena.}

\FullConference{The Golden Age of Cataclysmic Variables and Related Objects - VII (GOLDEN2025)\\
1-6 September, 2025\\
Mondello, Palermo, Italy\\}

\begin{document}
\maketitle

\section{The evolution of cataclysmic variables}\label{sect:CVevolution}

The long-term evolution of cataclysmic variables (CVs) is driven primarily by angular momentum losses, which cause the orbital separation between the white dwarf (WD) and the companion to shrink over time. As the orbital separation of the system decreases, the orbital period ($P_{\rm orb}$) also decreases, resulting in a relatively well-defined evolutionary track for CVs from longer to shorter orbital periods (Paczynski 1976 \cite{paczynski1976}; Kolb 1993 \cite{kolb1993}; Warner 2003 \cite{warner2003}; Knigge, Baraffe, and Patterson 2011 \cite{knigge2011}). This secular evolution is also reflected in the physical properties of the companion, which gradually transitions from a main-sequence star to a very late-type and eventually degenerate star (Giovannelli 2008 \cite{giovannelli2008}; Knigge, Baraffe, and Patterson 2011 \cite{knigge2011}). As the donor loses mass, the mass-transfer rate that it is able to sustain is also predicted to decline, with CVs evolving from high secular accretion rate systems (${\dot{M}}_{\rm acc}  \sim$ 10$^{-8}$ M$_{\odot}$/yr) to low secular accretion rate systems (${\dot{M}}_{\rm acc} \sim$ 10$^{-11}$ M$_{\odot}$/yr). 

The classical starting point of the evolution is a CV with relatively long orbital periods (P$_{\rm orb}>$ 6h) and companions masses of around M$_{\rm donor}\sim$0.7M$_{\odot}$. These systems evolve towards shorter periods until they reach the period gap (2h $<$ P$_{\rm orb}<\,$3h). Within the period gap, the standard model for the evolution of CVs predicts that magnetic braking as a mechanism for angular momentum loss becomes inefficient, causing the donor to contract within its Roche lobe and the system to detach (Kolb 1998 \cite{kolb1998}). The system evolves through the period gap as a detached binary, re-emerging at the lower edge of the gap as an active CV once the shrinking Roche lobe allows the donor to fill it again.

\begin{figure}[h]
  \begin{minipage}[c]{0.60\textwidth}
    \includegraphics[width=\textwidth,trim=0.9cm 0.6cm 1.9cm 1.9cm,clip]{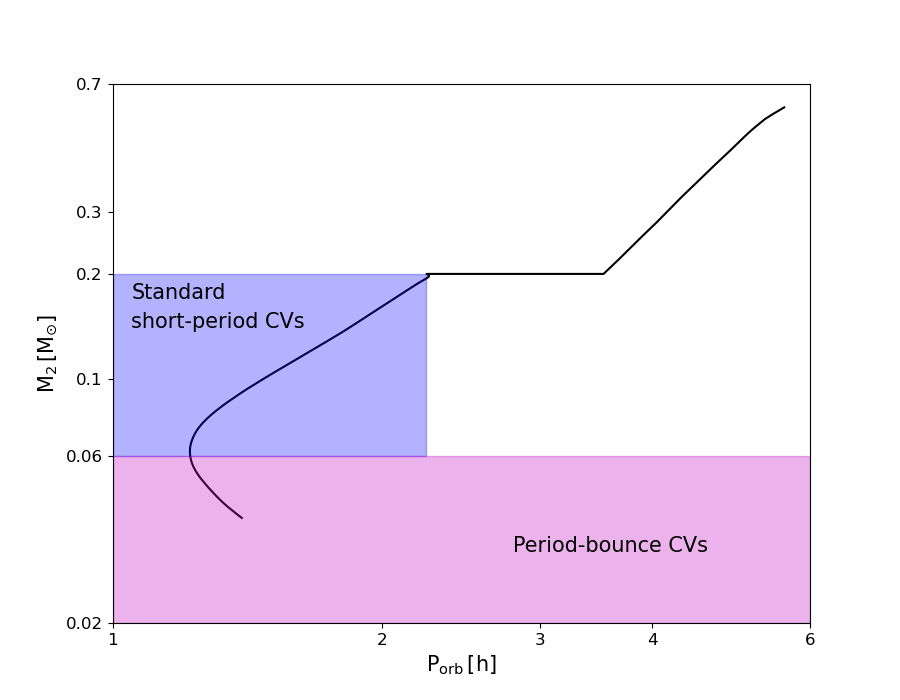}
  \end{minipage}\hfill
  \begin{minipage}[c]{0.39\textwidth}
    \caption{Classification of short-period CVs according to their evolutionary status into standard short period CVs (blue area) and period-bounce CVs (pink area). Standard evolution track for CVs from Knigge, Baraffe, and Patterson 2011 \cite{knigge2011} is shown in black as a reference. } \label{fig:CVevolution}
  \end{minipage}
\end{figure}

Angular momentum loss, mainly through gravitational radiation rather than magnetic braking, continues to shrink the orbital separation of the CV below the period gap. The CV continues its evolution towards even shorter periods until it reaches a period minimum located at $P_{\rm min} \approx 80$ \,min. At this stage, the donor mass has typically been reduced to 0.06M$_{\odot}$, placing it within the brown dwarf regime (see e.g., Howell, Nelson, and Rappaport \cite{howell2001}; Patterson 2011 \cite{patterson2011}; Pretorius, Knigge and Schwope 2013 \cite{pretorius2013}; Goliasch and Nelson 2015 \cite{goliasch}; McAllister et al. 2019 \cite{mcallister2019}; Pala et al. 2020 \cite{pala2020}; Pala et al. 2022 \cite{pala2022}). At the period minimum, the donor can no longer maintain thermal equilibrium because its mass-loss timescale has become significantly shorter than its Kelvin–Helmholtz timescale (King 1988 \cite{king1988}). At this stage, the donor has evolved into a very late-type object, effectively a brown dwarf unable to sustain ordinary hydrogen burning, having still the possibility of deuterium burning (Howell, Nelson, and Rappaport 2001 \cite{howell2001}). For such objects, the equation of state implies an inverse mass–radius relation ($R \propto M^{-1/3}$), such that continued mass loss causes the donor to expand rather than contract. This structural response drives an increase in the orbital separation, leading to a reversal of the secular evolution and causing the CV to bounce back toward longer orbital periods. CVs that have passed through this evolutionary turning point of "bouncing-back" to longer periods are known as period-bouncers (Patterson 1998 \cite{patterson1998}).

Fig.~\ref{fig:CVevolution} illustrates this evolutionary sequence by distinguishing between the two principal populations of short-period CVs: (1) standard short-period CVs, evolving toward the period minimum with main-sequence companions, and (2) period-bounce CVs, which have evolved past the period minimum and contain degenerate companions. 

In the following sections, we discuss how the capabilities of {\it XMM-Newton} and eROSITA facilitate the detection, characterization and identification of standard short-period CVs and period-bouncers, with particular emphasis on magnetic systems.

\section{Cataclysmic variables in XMM-Newton}\label{sect:xmm}

\begin{figure}
\begin{center}
  \includegraphics[width=0.52\columnwidth,trim=6.6cm 5.5cm 6.2cm 0.6cm,clip]{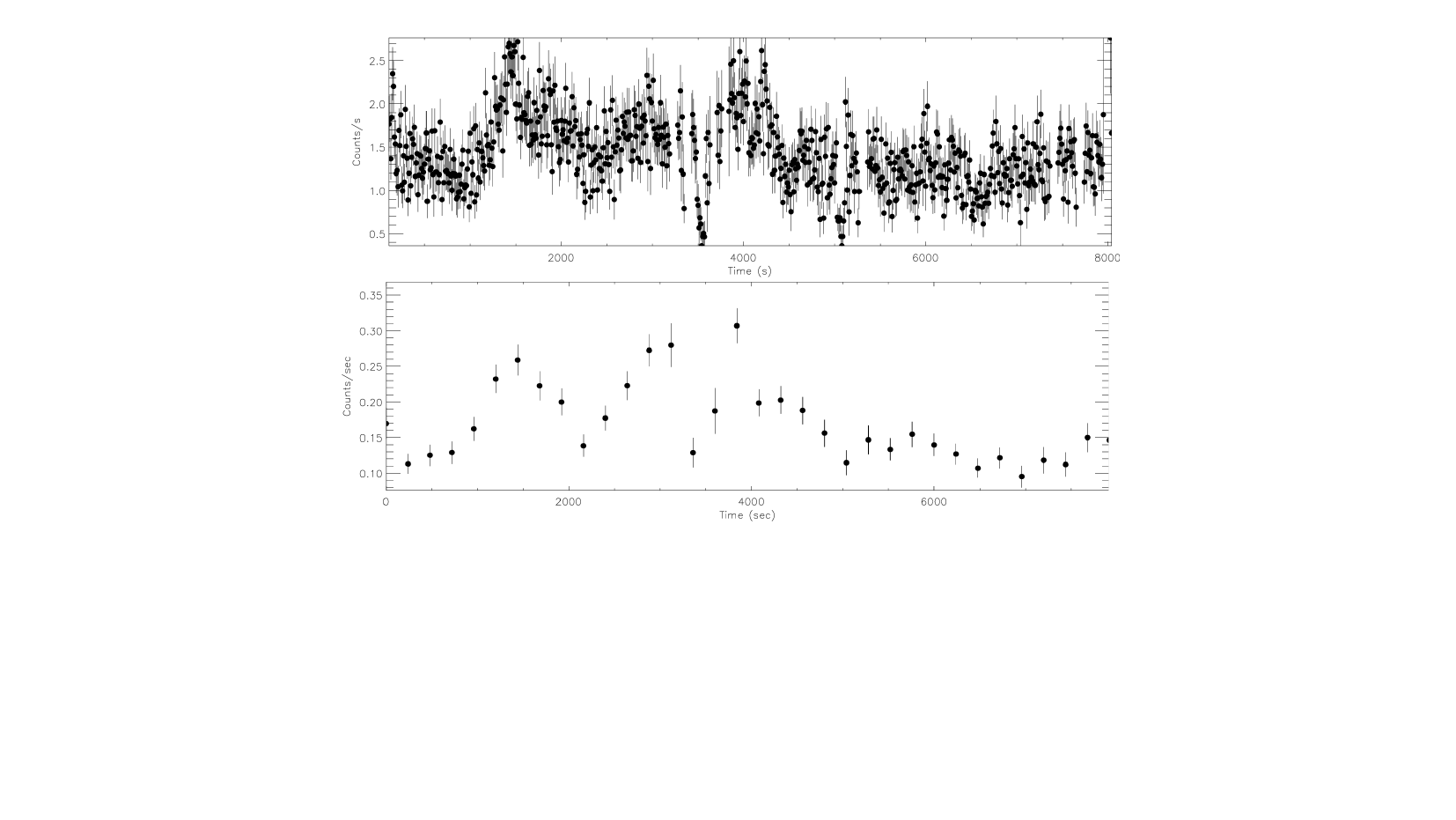}
  \includegraphics[width=0.47\columnwidth,trim=3.6cm 2.7cm 7.8cm 1.6cm,clip]{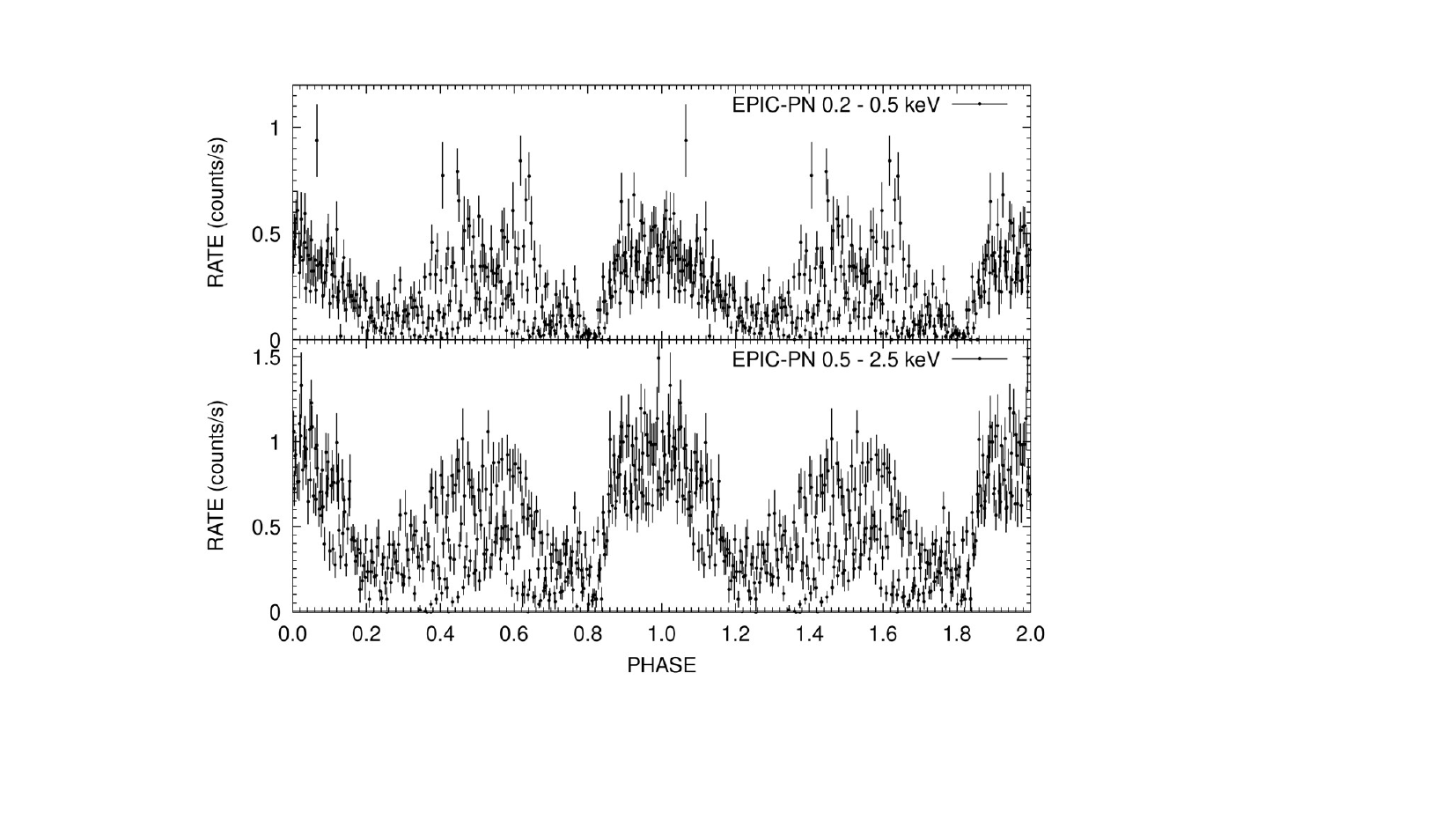}
  \caption {Examples of \textit{XMM-Newton} light curves of CVs. \textit{Left panel:} Confirmed period-bouncer WZ Sge, a non-magnetic CV with a light curved binned using 10s (upper panel) and 240s (bottom panel) (Nucita et al. 2014 \cite{nucita2014}). \textit{Right panel:} V496 UMa, a polar with two-pole accretion geometry observable in both soft (0.2-0.5keV) and medium (0.5-2.5keV) X-ray bands (Ok and Schwope 2022 \cite{ok2022}). }
  \label{fig:xrayLigthcurve}
 \end{center}
 \end{figure}

\begin{figure}[h]
  \begin{minipage}[c]{0.53\textwidth}
    \includegraphics[width=\textwidth,trim=0.1cm 0.3cm 0.1cm 0.55cm,clip]{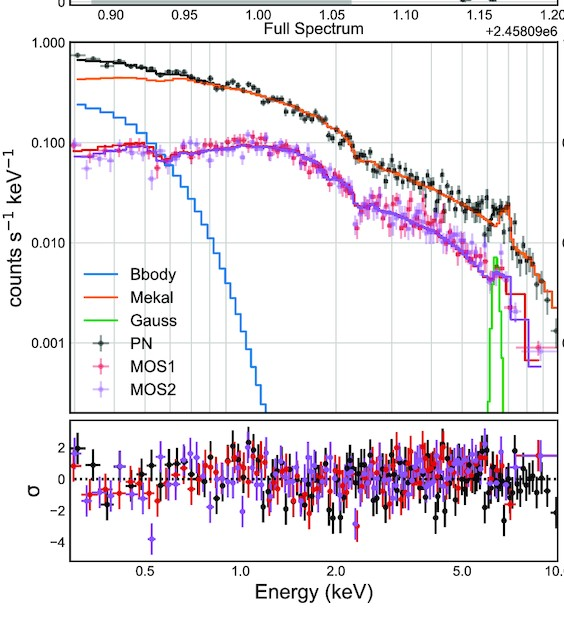}
  \end{minipage}\hfill
  \begin{minipage}[c]{0.46\textwidth}
    \caption{\textit{XMM-Newton} spectrum of V496 UMa, a magnetic CV, with a best fit model (in black) obtained using \texttt{XSPEC} (Arnaud 1996 \cite{arnaud1996}) made up of three components: blackbody model (Bbody), hot diffuse gas model with line emissions from several elements (mekal) and Gaussian line profile (Gauss). Adapted from Kennedy, Littlefield and Garnavich 2022 \cite{kennedy2022}.} \label{fig:xraySpectrum}
  \end{minipage}
\end{figure}

The {\it XMM-Newton} space observatory, launched on 10 December 1999 by the European Space Agency (ESA) as part of the Horizon 2000 Science Programme, remains one of the most powerful X-ray observatories for studying compact binaries such as CVs. {\it XMM-Newton} typically observes a fixed position on the sky for exposures of at least $5$\,ks, often significantly longer. These long, uninterrupted pointings yield high-cadence X-ray light curves, capable of monitoring short term X-ray variability and modulation (see Fig.~\ref{fig:xrayLigthcurve}), and high signal-to-noise spectra, that allow the study of multiple components contributing to the X-ray emission (see Fig.~\ref{fig:xraySpectrum}).

The high-cadence {\it XMM-Newton} light curves provided by the EPIC instruments (pn, MOS1, and MOS2) allow the detection of variability and modulations often associated with the orbital period, the WD spin, pulsations, or accretion-related instabilities. Because {\it XMM-Newton} provides continuous coverage typically for durations longer than the CV variability timescales (of few hours), it is particularly well suited for characterizing multi-periodic behavior that may be difficult to detect with other telescopes or surveys.

The {\it XMM-Newton} light curves of WZ Sge and V496 UMa are shown in Fig.~\ref{fig:xrayLigthcurve} to exemplify different types of variability that can be observed in X-rays for CVs. WZ Sge, a non-magnetic period-bouncer, displays variability that has been associated with two periodicities: one related to the spin period or pulsations from the WD, and another one related to the orbital period of the system (Nucita et al. 2014 \cite{nucita2014}). V496 UMa, a magnetic CV, shows strong modulations at both the orbital period and half the orbital period, signatures of two-pole accretion geometry (Ok and Schwope 2022 \cite{ok2022}), with a slight difference in the modulations presented by the soft and medium X-ray bands, providing insight into the spatial distribution of the accretion regions and their spectral hardness.

Beyond time-domain studies, {\it XMM-Newton}’s broad-band spectroscopy (0.2–10 keV) is a powerful diagnostic of the accretion environment in CVs. In Fig.~\ref{fig:xraySpectrum} the {\it XMM-Newton} spectrum of V496 UMa has been fitted with a composite model that captures the main physical components of the accretion process (Kennedy, Littlefield and Garnavich 2022 \cite{kennedy2022}). (1) A blackbody model, representing the soft X-ray emission typically associated with reprocessed radiation from the WD surface. In magnetic systems, material is being funneled along magnetic field lines forming an accretion column. Post-shock plasma from this accretion column radiates hard X-rays, some of which are absorbed and re-emitted by the WD surface as soft blackbody-like emission (Lamb and Masters 1979 \cite{lamb1979}). (2) A hot diffuse gas model with line emission from highly ionized elements, tracing the medium and hard X-ray radiation arising from the post-shock plasma in the accretion column and its interaction with the WD surface (Mukai 2017 \cite{mukai2017}). (3) a Gaussian line profile used to reproduce the iron emission line at $\sim$6.5 keV generated possibly due to fluorescence caused by the reflection of hard X-rays off the WD surface (Rothschild et al. 1981 \cite{rothschild1981})

\subsection{Detection of polars} \label{subsect:XMMpolars}

One of the most interesting types of CVs to study with {\it XMM-Newton} are polars. These short-period CVs contain strongly magnetic WDs with field strengths $B_{WD}\,\gtrsim 10$MG. In these systems, the formation of an accretion disk is prevented, as the Alfv\'en radius is greater than the circularization radius. Accretion happens instead along accretion columns which are formed as matter is captured and transported through the magnetic field lines of the WD (Ulla, Giovannelli and Martinez-Pais 1999 \cite{ulla1999}). Once the accretion column impacts the surface of the WD, at the location of the magnetic poles, it generates an X-ray bright hot-spot. 

Polars, contrary to intermediate polars, exhibit some of the shortest orbital periods among magnetic CVs and are therefore found near the end of their secular evolution. The strong magnetic field of the WD as well as the long evolution time of polars, required to reach their evolutionary stage as short-period CVs, results in the observed synchronization of these systems, with the rotation of the WD being synchronous to the orbital period of the binary (P$_{\textrm{spin}} =$ P$_{\textrm{orb}}$; Cropper 1990 \cite{cropper1990}).

\begin{figure}[h]
  \begin{minipage}[c]{0.63\textwidth}
    \includegraphics[width=\textwidth,trim=17.5cm 1.3cm 2.5cm 10.7cm,clip]{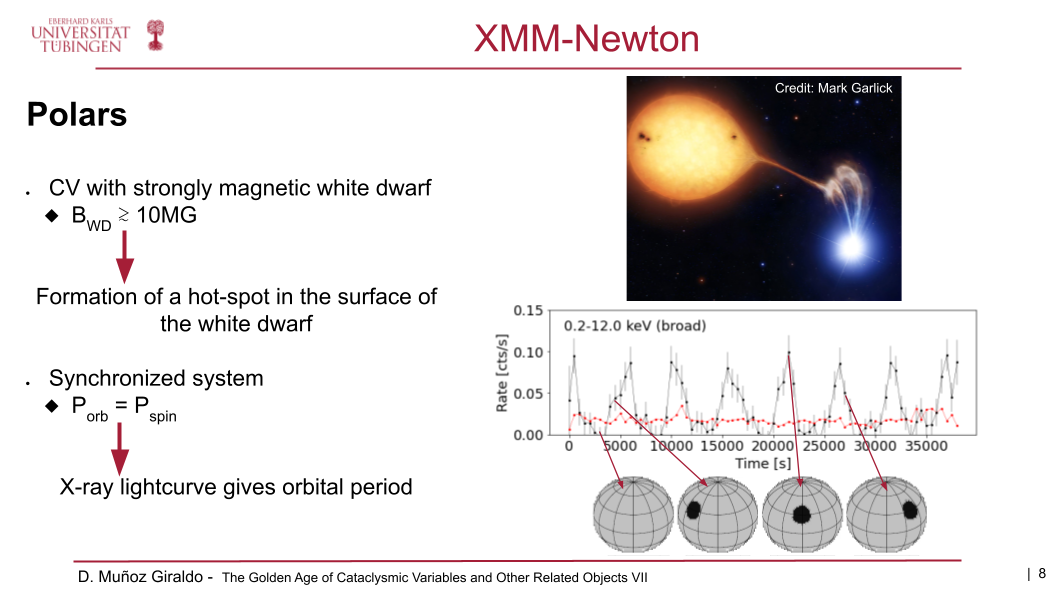}
  \end{minipage}\hfill
  \begin{minipage}[c]{0.36\textwidth}
    \caption{\textit{XMM-Newton} light curve of period-bounce polar SDSS J151415.65+074446.4 used to show the expected modulation for polars where the rotation of the WD is synchronized with the orbital period of the system. Adapted from Muñoz-Giraldo et al. 2023 \cite{munoz2023}. } \label{fig:polarSynchronization}
  \end{minipage}
\end{figure}

Because polars lack an accretion disk and are characterized by having synchronized orbital-spin periods, their X-ray light curves exhibit strong orbital modulations that closely trace the binary orbital period. These modulations arise from the changing visibility of one or both magnetic poles of the WD as it rotates. Fig.~\ref{fig:polarSynchronization} presents a schematic representation of one-pole accretion. When the primary hot-spot rotates out of the field of view, the light curve exhibits a minimum count rate (often consistent with zero). As the hot-spot rotates back into the field of view the count rate increases, reaching a maximum value when the hot-spot is entirely visible. Therefore, the time between successive maxima (or minima) represents the orbital period of the system in the case of one-pole accretion.

In contrast, when accretion occurrs onto both magnetic poles of the WD, the X-ray light curve will display two maxima per orbital cycle, as two different X-ray emitting regions are coming into the field of view for every rotation of the WD. The associated two periodicities correspond to the orbital period of the system (given by the time between maxima at similar count rates) and half of the orbital period (given by the time between any consecutive maxima or minima). Because accretion will still happen preferentially towards one of the poles of the WD, the two maxima often have different amplitudes, allowing the primary and secondary hot-spots to be distinguished.

\begin{figure}
\begin{center}
  \includegraphics[width=\columnwidth,trim=0.2cm 5.9cm 0.3cm 2.5cm,clip]{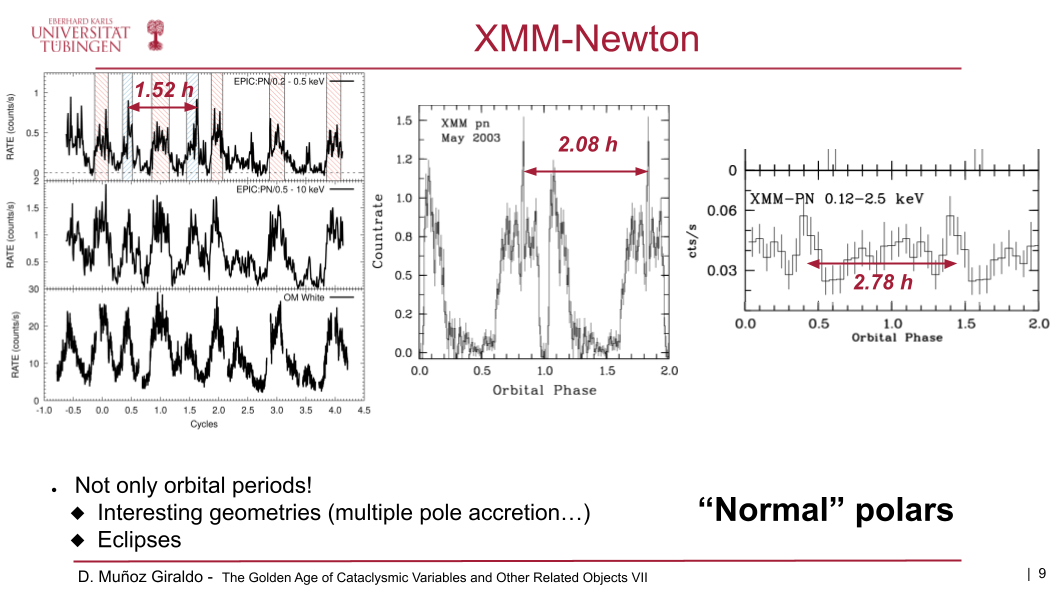}
  \caption {\textit{XMM-Newton} light curves of polars. \textit{Left panel:} V496 UMa, a polar with two-pole accretion geometry observable in both soft (0.2-0.5keV) and hard (0.5-10keV) X-ray bands as well as in the optical band. Emission from each WD pole is highlighted by a different color (adapted from Ok and Schwope 2022 \cite{ok2022}). \textit{Middle panel:} HU Aqr, a polar with one-pole accretion exhibiting an eclipse (adapted from Schwarz et al. 2009 \cite{schwarz2009}). \textit{Right panel:} WX LMi, a polar with two-pole accretion geometry observable in the soft (0.12-2.5keV) X-ray band (adapted from Vogel, Schwope, and Gänsicke 2007 \cite{vogel2007}). }
  \label{fig:xrayLigthcurvePolars}
 \end{center}
 \end{figure}

The {\it XMM-Newton} light curves of V496 UMa, HU Aqr and WX LMi are shown in Fig.~\ref{fig:xrayLigthcurvePolars} to illustrate the differences between one-pole and two-pole accretion. V496 UMa and WX LMi are polars presenting two-pole accretion (Ok and Schwope 2022 \cite{ok2022}; Vogel, Schwope, and Gänsicke 2007 \cite{vogel2007}), which can be identified due to the presence of two different maxima for each orbital cycle (in the light curve of V496 UMa the maxima corresponding to the primary and secondary hot-spots are differentiated using colors). HU Aqr, a polar, displays variability arising from accretion through a single pole (Schwarz et al. 2009 \cite{schwarz2009}) as well as the presence of a deep eclipse, which provides further constraints on the geometry of the system. The eclipses and modulations shown by the {\it XMM-Newton} light curves of polars can be used to infer key parameters such as the binary inclination, which is critical for determining the mass of the components and the mass accretion rate of the system (Schwope et al. 2002 \cite{schwope2002}; Schwarz et al. 2009 \cite{schwarz2009};  Nucita et al. 2011 \cite{nucita2011}).

Another interesting characteristic of polars is their tendency to switch between high and low mass accretion rate states. These transitions are thought to arise from changes in the mass accretion rate from the donor star, possibly driven by star spots drifting in front the inner Lagrangian point (Livio and Pringle 1994 \cite{livio1994}, Kafka and Honeycutt \cite{kafka2005}). Polars can spend a substantial fraction of their lifetimes in low states, with AM Her (the prototypical polar) spending roughly half of its time in a low state (Hessman, Gänsicke, and Mattei 2000 \cite{hessman2000}). This behavior not only complicates their identification in large surveys but also introduces biases in population studies, as many systems may be detected (or missed) depending on their accretion state (Pretorius, Knigge, and Schwope 2013 \cite{pretorius2013}). 

\begin{figure}
\begin{center}
  \includegraphics[width=\columnwidth,trim=0.5cm 1.3cm 0cm 6.9cm,clip]{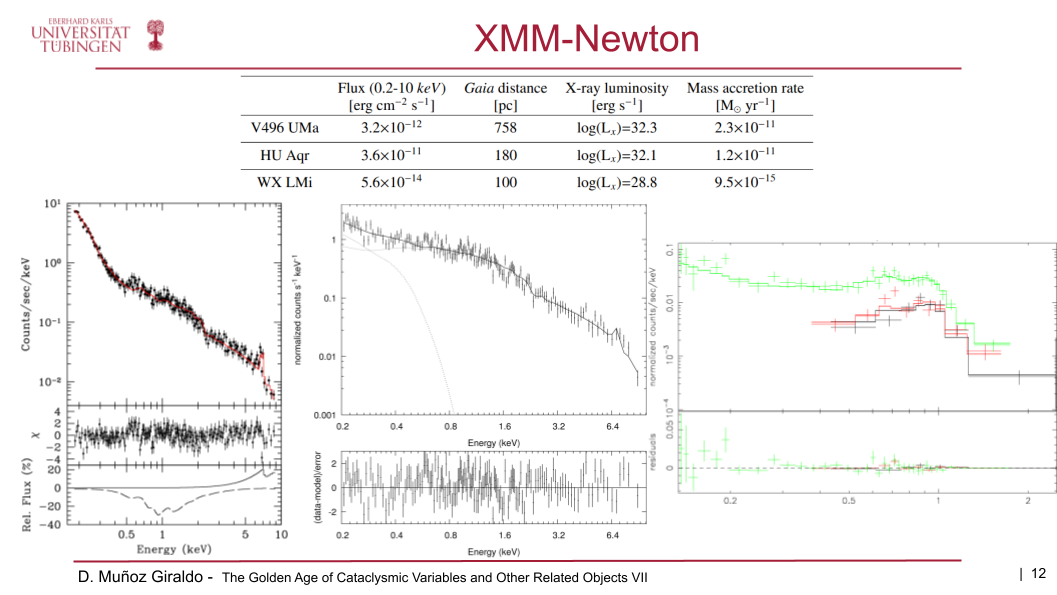}
  \caption {\textit{XMM-Newton} spectra of polars. \textit{Left panel:} V496 UMa observed during a high state (Ok and Schwope 2022 \cite{ok2022}). \textit{Middle panel:} HU Aqr observed during a high state (Schwarz et al. 2009 \cite{schwarz2009}). \textit{Right panel:} WX LMi observed during a low state (Vogel, Schwope, and Gänsicke 2007 \cite{vogel2007}). }
  \label{fig:xraySpectraPolars}
 \end{center}
 \end{figure}

 The {\it XMM-Newton} spectra of V496 UMa and HU Aqr showed in Fig.~\ref{fig:xraySpectraPolars} were taken during a high state (Ok and Schwope 2022 \cite{ok2022}; Schwarz et al. 2009 \cite{schwarz2009}), which is associated with elevated observed X-ray fluxes ($\sim$3$\times 10^{-12}$ erg cm$^{-2}$ s$^{-1}$ and $\sim$4$\times 10^{-11}$ erg cm$^{-2}$ s$^{-1}$ respectively) and relatively high instantaneous mass accretion rates ($\sim$2$\times 10^{-11}$ M$_{\odot}$ yr$^{-1}$ and $\sim$1$\times 10^{-11}$ M$_{\odot}$ yr$^{-1}$ respectively). In these high states, the accretion column is fully sustained by high instantaneous mass accretion rates, which produces substantial medium and hard X-ray emission, with detectable flux beyond 5 keV. Both systems show a fluorescence emission feature at $\sim$6.5 keV, consistent with reflection off the WD surface. 
 
 On the other hand, the {\it XMM-Newton} spectrum of WX LMi was taken during a low state (Vogel, Schwope, and Gänsicke 2007 \cite{vogel2007}), associated with a much lower observed X-ray flux ($\sim$6$\times 10^{-14}$ erg cm$^{-2}$ s$^{-1}$) and very low instantaneous mass accretion rate ($\sim$9$\times 10^{-15}$ M$_{\odot}$ yr$^{-1}$). In this low state, the accretion column is weak or absent, and the spectrum is dominated by soft X-ray emission from the heated WD surface. Low state spectra are typical of highly evolved CVs, including period-bouncers or other CVs around the period minimum.

Spectral fitting of polars provides direct insight into the physical mechanisms responsible for their X-ray emission. The medium and hard X-ray components trace the temperature and structure of the post-shock plasma, while the soft component reveals the extent of re-processing on the WD surface. Together, these diagnostics constrain instantaneous mass accretion rates, and the geometry of the magnetic field and hot-spot.

\subsubsection{Detection of period-bounce polars}

Among the known population of polars, a small but growing subset has been identified as period-bounce polars, having evolved past the period minimum and now moving toward longer periods. At this evolutionary stage, and mainly due to the very low mass of the donor (which is now a brown dwarf), the system exhibits very low mass accretion rates. As a result, period-bounce polars are no longer able to maintain high states and are instead persistently observed in low states, characterized by faint X-ray and optical fluxes (Muñoz-Giraldo et al. 2023 \cite{munoz2023}). This behavior makes them more difficult to detect in traditional surveys and introduces observational biases against identifying period-bounce polars.

\begin{figure}[h]
  \begin{minipage}[c]{0.63\textwidth}
    \includegraphics[width=\textwidth,trim=0.1cm 0.1cm 0.1cm 0.1cm,clip]{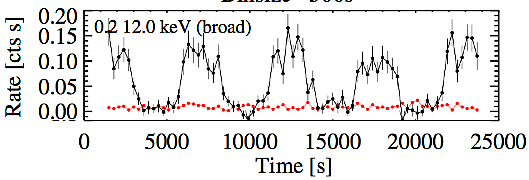}
  \end{minipage}\hfill
  \begin{minipage}[c]{0.36\textwidth}
    \caption{\textit{XMM-Newton} light curve of period-bounce polar V379 Vir (Stelzer et al. 2017 \cite{stelzer2017}). } \label{fig:xrayLigthcurveSDSS1212}
  \end{minipage}
\end{figure}

However, despite their low mass accretion rates, they still present the typical characteristics of polars. Two confirmed period-bounce polars, SDSS J151415.65+074446.4 and V379 Vir, both present in their {\it XMM-Newton} light curves the modulation associated with emission arising from one-pole accretion (see Fig.~\ref{fig:polarSynchronization} and Fig.~\ref{fig:xrayLigthcurveSDSS1212}). These modulations were used to determine their orbital periods (87.9min and 88.3min respectively) and constrain the geometry of the system (Stelzer et al. 2017 \cite{stelzer2017}; Muñoz-Giraldo et al. 2023 \cite{munoz2023}). The detection of the modulation, even at low mass accretion rates, highlights the importance of the analysis of {\it XMM-Newton} light curves in order to identify magnetic CVs close to or beyond the period minimum.

\begin{figure}
\begin{center}
  \includegraphics[width=\columnwidth,trim=0.3cm 1.3cm 0.3cm 6.4cm,clip]{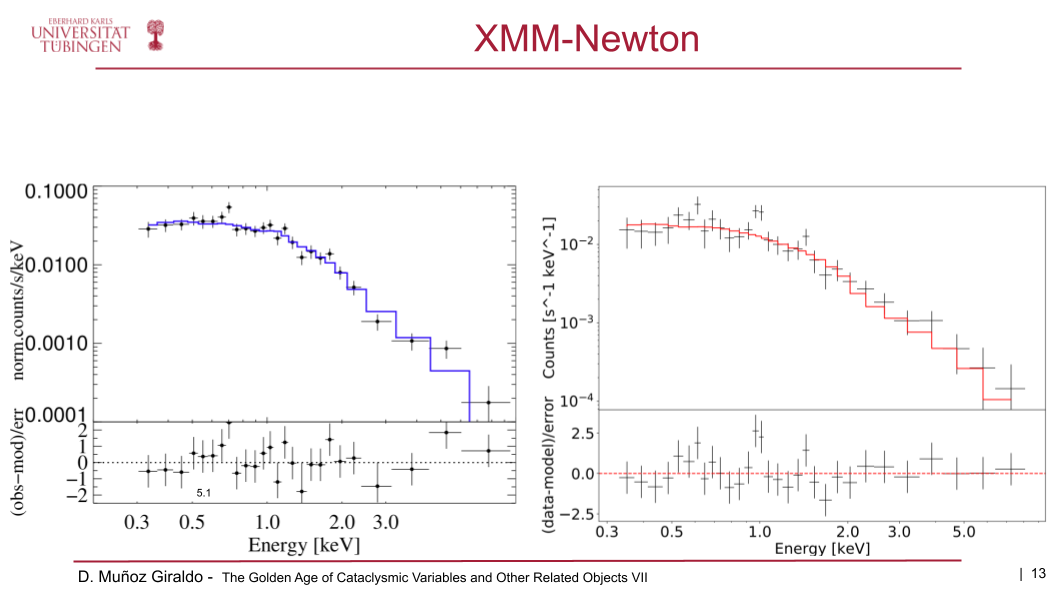}
  \caption {\textit{XMM-Newton} spectra of period-bounce polars. \textit{Left panel:} V379 Vir (Stelzer et al. 2017 \cite{stelzer2017}). \textit{Right panel:} SDSS J151415.65+074446.4 (Muñoz-Giraldo et al. 2023 \cite{munoz2023}). }
  \label{fig:xraySpectraPBs}
 \end{center}
 \end{figure}

As expected from polars in a persistent low state, the {\it XMM-Newton} spectra of these systems (Fig.~\ref{fig:xraySpectraPBs}) are characterized by being soft, with the majority of photons detected below 5keV. Both V379 Vir and SDSS J151415.65+074446.4 exhibit very low observed X-ray fluxes ($\sim$1$\times 10^{-13}$ erg cm$^{-2}$ s$^{-1}$ and $\sim$7$\times 10^{-14}$ erg cm$^{-2}$ s$^{-1}$ respectively) and very low instantaneous mass accretion rates ($\sim$4$\times 10^{-14}$ M$_{\odot}$ yr$^{-1}$ and $\sim$3$\times 10^{-14}$ M$_{\odot}$ yr$^{-1}$ respectively). These values place period-bounce polars among the X-ray faintest known magnetic CVs, highlighting the need for deep, sensitive observations to uncover more members of this population.

The X-ray timing and spectral analysis enabled by {\it XMM-Newton} allows for the identification and characterization of magnetic CVs across a broad range of evolutionary stages. From high state polars to faint period-bounce CVs, {\it XMM-Newton} observations allows to: determine accretion geometries, measure instantaneous mass accretion rates, and detect modulations associated to orbital or rotational motions. These capabilities are essential for completing the census of highly evolved CVs and for testing the prediction of population synthesis models.

\section{Cataclysmic variables in eROSITA}\label{sect:eROSITA}

The eROSITA telescope was launched in 2019 onboard the Russian-German "Spectrum-Roentgen-Gamma" (SRG) observatory as part of a joint mission by Germany and Russia. Differently to {\it XMM-Newton}, which performs targeted pointed observations, eROSITA was designed as an all-sky survey, scanning the entire sky every six months. Between launch and the mission interruption in February 2022, eROSITA completed four all-sky surveys (eRASS 1 to 4) and was only able to partially finish the fifth one before the telescope was shut down indefinitely. Data rights for eROSITA are divided between the partner nations, with the German eROSITA consortium having exclusive access to the western Galactic hemisphere (Galactic longitude $l \geq 180^\circ$), which we refer to throughout as the German eROSITA sky. 

\begin{figure}[h]
  \begin{minipage}[c]{0.67\textwidth}
    \includegraphics[width=\textwidth,trim=0.1cm 0.1cm 0.1cm 0.1cm,clip]{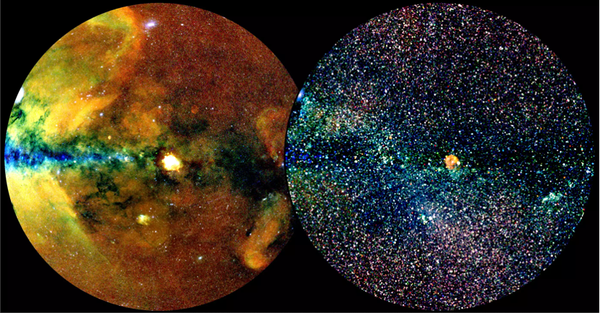}
  \end{minipage}\hfill
  \begin{minipage}[c]{0.32\textwidth}
    \caption{The eROSITA X-ray sky showing extended sources (left) and detected point sources (right). {\it Image credit:} MPE / J. Sanders for the eROSITA Consortium.} \label{fig:eROSITApointSources}
  \end{minipage}
\end{figure}

The first public data release of eROSITA was officially released on January 31st 2024, containing data from 930000 point sources in the soft band over the German eROSITA sky (Fig.~\ref{fig:eROSITApointSources}). This represents an increase of more than 60$\%$ in the number of soft X-ray sources compared to previous catalogs (Merloni et al. 2024 \cite{merloni2024}), underscoring the huge impact of the eROSITA survey for X-ray studies.

Although eROSITA concluded earlier than originally envisioned, therefore not being able to complete the eight all-sky surveys that were initially planned, the combined data from the first four all-sky surveys (eRASS:4) still represents the most comprehensive soft X-ray view of the sky to date, containing $\sim$3 million point sources. This very large amount of detected sources is ideal to conduct population studies, as it allows to study the general characteristics of different source types, helping distinguish between them using diagnostic diagrams. 

For CVs in particular, eROSITA data can be applied to construct typical X-ray diagnostic diagrams, with the two more commonly used being distance versus X-ray luminosity and {\it Gaia} colors versus X-ray-to-optical flux ratio. These diagnostic diagrams are now central to the identification of new CV candidates in the eROSITA catalog, including highly evolved systems with intrinsically low luminosities like period-bouncers (Muñoz-Giraldo, Stelzer and Schwope 2024 \cite{munoz2024a}; Schwope et al. 2024a \cite{schwope2024a}; Schwope et al. 2024b \cite{schwope2024b}).

\subsection{What are the goals?}\label{subsect:goals}

With the immense population of X-ray sources that eROSITA has uncovered, one of the primary aims of ongoing work is to address long standing open questions regarding the nature and evolution of CVs. eROSITA provides the statistical power needed to study CVs as a population rather than as a collection of individually discovered systems. Our scientific objectives focus on two broad themes: binary evolution and the Galactic Ridge X-ray Emission (GRXE).

From an evolutionary perspective, eROSITA offers the opportunity to validate the predictions made by population synthesis studies using observational data. This includes quantifying the relative numbers of short-period CVs, identifying highly evolved systems such as period-bouncers, assessing the fraction of magnetic versus non-magnetic CVs across the Galaxy, establishing the role of CVs as potential progenitors of Type Ia supernovae, and determining the contribution of CVs to the Galactic gravitational-wave foreground. The discovery potential for intrinsically faint systems is especially relevant, as their underrepresentation in previous surveys has been a major obstacle to understanding the full evolution of CVs.

The second major goal concerns the contribution of CVs, particularly magnetic systems, to the GRXE. The integrated X-ray emission of the Galactic plane has long been suspected to originate from a large population of faint, unresolved point sources, among which CVs (and active binaries) are expected to play a dominant role. The all-sky coverage of eROSITA will enable a systematic assessment of the X-ray luminosity function of CVs and their spatial distribution within the Galaxy, both of which are essential for quantifying their contribution to the GRXE.

Together, these goals highlight the important role that eROSITA plays in shaping the next generation of population studies of CVs.


\subsubsection{Evolution}

The evolution of CVs, from long-period systems with relatively massive donors and high mass accretion to short-period systems characterized by low mass accretion rates and degenerate donors, has been well established observationally, with systems populating each stage in the CV evolution. However, there is a wide range of theoretical models attempting to reproduce the observed properties of this evolution (Politano 1996 \cite{politano1996}; Zorotovic, Schreiber and Gänsicke 2011 \cite{zorotovic2011}; Wijnen, Zorotovic and Schreiber 2015 \cite{wijnen2015}; Schreiber, Zorotovic and Wijnen 2016 \cite{schreiber2016}; Belloni et al. 2020 \cite{belloni2020}; Schreiber and Belloni 2025 \cite{schreiber2025}), each with distinct strengths and limitations. To date, none of the models reproduces all aspects of the observed evolution of CVs, namely the period distribution, the fraction of magnetic CVs, the mass accretion rates, or the abundance of evolved systems. The unprecedented sample size that is expected to be provided by eROSITA offers the opportunity to address the areas of CV evolution that are still unexplained by constructing statistically robust, volume-limited, and less biased CV samples, thereby informing the next generation of evolutionary models.

\begin{itemize}
    \item {\it Magnetism in CVs:} The magnetic field strength of the WD, an important ingredient when describing the history of angular momentum loss and therefore the evolution of CVs, is either not included or treated in an overly simplified way in most evolutionary models. This is important for the study of CVs when considering that one of the two mechanism of angular momentum loss that governs their evolution is magnetic wind braking, which arises from the stellar wind associated with the donors's magnetic activity (see e.g Mestel 1968 \cite{mestel1968}, Verbunt and Zwaan 1981 \cite{verbunt1981}) ejecting material and angular momentum from the system. Magnetic wind braking is expected to be reduced in magnetic CVs, as the wind from the donor may be trapped within the magnetosphere of the WD, causing a difference in the rate of angular momentum loss between magnetic and non-magnetic CVs (Webbink and Wickramasinghe 2002 \cite{webbink2002}). However, it is still to be explored to which extent the magnetism of the WD actually affects the efficiency of magnetic wind braking as a mechanism for angular momentum loss and its implications for the orbital period distribution. Additionally, the fraction of WDs exhibiting strong magnetic fields is significantly higher in CVs ($\sim$30$\%$) than among single WDs (2$\%$ to 20$\%$; Liebert et al. 2003 \cite{liebert2003}; Kawka et al. 2007 \cite{kawka2007}; Giammichele, Bergeron and Dufour 2012 \cite{giammichele2012}; Kepler et al. 2013 \cite{kepler2013}; Sion et al. 2014 \cite{sion2014}; Kepler et al. 2015 \cite{kepler2015}; Pala et al. 2020 \cite{pala2020}; Rodriguez et al. 2025 \cite{rodriguez2025}). The origin of this discrepancy is unknown. The eROSITA CV sample will provide the first opportunity to systematically test the role of magnetism across the entire population.
    \item {\it CVs as a channel for the formation of Type-Ia supernovae:} Accreting WDs in CVs have long been considered potential progenitors of Type-Ia supernovae via the single-degenerate channel (Kafka 2012 \cite{kafka2012}). While classical CVs generally exhibit mass loss through nova eruptions, some subclasses of CVs, particularly systems with high mass accretion rates, may allow the WD to experience net mass growth (Maoz, Mannucci, and Nelemans 2014 \cite{maoz2014}). However, the frequency of such systems, their mass-retention efficiencies, and their long-term evolutionary pathways remain poorly constrained observationally. eROSITA, by expanding the known population of both high mass accretion rate systems and evolved short-period systems, will allow improved estimates of how many CVs host massive WDs, how many CVs exhibit sustained accretion capable of increasing WD mass, and whether these systems could contribute meaningfully to the Type-Ia supernova rate.
    \item {\it Contribution of CVs to the gravitational wave foreground:} Short-period CVs, particularly those near or past the period minimum, are expected to be persistent sources of low-frequency gravitational waves detectable by future missions such as {\it LISA} (Scaringi et al. 2023 \cite{scaringi2023}). Interacting systems are predicted to be detected amongst the Galactic population of compact binaries producing the gravitational wave foreground, yet current estimates remain uncertain due to the incomplete census of faint, evolved CVs (Kupfer et al. 2018 \cite{kupfer2018}). eROSITA’s sensitivity to low-luminosity systems, including magnetic and non-magnetic period-bouncers, will enable more accurate measurements of space densities and orbital period distributions, allowing the refining of predictions for the Galactic gravitational wave foreground and identifying individual binaries that could represent verification sources for {\it LISA}.
    \item {\it The period-bouncer problem:} One of the most contentious disagreement between theory and observations concerns the predicted abundance of period-bouncers. A key prediction of theoretical models describing the evolution of CVs is that period-bouncers are expected to be the majority of the population, with estimates ranging between 40$\%$ and 80$\%$ (see e.g., Kolb 1993 \cite{kolb1993}; Goliasch and Nelson 2015 \cite{goliasch}; Schreiber, Zorotovic and Wijnen 2016 \cite{schreiber2016}; Belloni et al. 2020 \cite{belloni2020}). Observationally, however, only 3$\%$ to 25$\%$ of CVs have been identified as period-bouncers to date (Pala et al. 2020 \cite{pala2020}; Inight et al. 2023b \cite{inight2023b}; Rodriguez et al. 2025 \cite{rodriguez2025}; Muñoz-Giraldo et al. 2025 \cite{munoz2025}). Period-bouncers are intrinsically faint, evolved CVs with very low luminosities and mass transfer rates, making their identification challenging and possibly leading to selection biases against them that explain this under-representation (Patterson 2011 \cite{patterson2011}). Another recent explanation for the lack of observed period-bouncers suggests that a significant number of them might become detached for several gigayears due to the late appearance of strong WD magnetic fields (Schreiber and Belloni 2025 \cite{schreiber2025}). These detached period-bouncers would be fainter than standard period-bouncers due to their almost imperceptible mass accretion rates, making their observation and identification even more challenging. With its ability to detect faint X-ray sources across the entire sky, eROSITA is uniquely positioned to uncover both accreting and detached period-bouncers, thereby providing a decisive test of current evolutionary models.    
\end{itemize}

\subsubsection{Galactic Ridge X-ray Emission} \label{subsect:GRXE}
The Galactic Ridge X-ray Emission (GRXE) was first detected as an extended, apparently diffuse X-ray glow along the inner Galaxy (Worrall et al. 1982 \cite{worrall1982}; Warwick et al. 1985 \cite{warwick1985}). Although early observations suggested that the GRXE might arise from diffuse plasma (Koyama et al. 1986 \cite{koyama1986}), more recent studies, thanks to a set of deep {\it Chandra} observations targeting the inner Galaxy, showed that a substantial fraction of this emission is produced by the integrated light of numerous faint, unresolved point sources (Revnivtsev et al. 2006 \cite{revnivtsev2006}; Revnivtsev et al. 2009 \cite{revnivtsev2009}). In particular, intermediate polars are expected to dominate this emission due to their relatively hard spectra, high X-ray luminosities  and strong Fe-line complexes (Sazonov et al. 2006 \cite{sazonov2006}; Yuasa, Makishima and Nakazawa 2012 \cite{yuasa2012};  Warwick 2014 \cite{warwick2014}; Hailey et al. 2016 \cite{hailey2016}). However, the exact contribution of CVs to the GRXE remains difficult to quantify because current estimates of their Galactic space density, luminosity function, and fractions of different CV types are affected by selection biases.

eROSITA observations are expected to lead to the identification of a large proportion of all X-ray emitting CVs within 500pc (including both magnetic and non-magnetic CVs), as well period-bouncers within 250pc. These capabilities allow the construction of complete volume-limited samples that are highly complete within the corresponding distance limits. Such unbiased samples are crucial for determining the true space density of CVs, mapping their luminosity distributions, and quantifying the relative numbers of magnetic versus non-magnetic systems (Pretorius and Knigge 2012 \cite{pretorius2012}; Pala et al. 2020 \cite{pala2020}). Once these volume-limited samples are completed, it becomes possible to perform the first reliable estimate of the cumulative X-ray emission of the local CV population and compare it against the emission from the GRXE (Revnivtsev et al. 2009 \cite{revnivtsev2009}). In particular, eROSITA's high statistics will allow us to test whether intermediate polars alone can reproduce the observed hard GRXE spectrum, or whether a substantial contribution from non-magnetic systems is required.

Beyond clarifying the nature of the GRXE, the eROSITA CV population will also serve as a crucial input for models linking the GRXE to stellar population demographics and Galactic structure. By directly measuring the distribution of CVs as a function of Galactic latitude, distance, and scale height, eROSITA will provide the empirical foundation needed to connect individual binary evolution channels with the integrated X-ray background of the Milky Way.

\subsection{Populations} \label{subsection:populations}

In order to address the questions regarding CV evolution and the GRXE presented in Sect.~\ref{subsect:goals}, a complete and unbiased census of CVs is required. The eRASS:4 survey offers precisely this opportunity, enabling for the first time, a large-scale characterization of the Galactic CV population in X-rays. In this following section, we examine the properties and distribution of the known CVs already identified in the eROSITA catalog and outline how these detections will be used in future population studies.

\begin{figure}
\begin{center}
 \includegraphics[width=0.45\columnwidth,trim=21.1cm 0cm 2.4cm 0cm,clip]{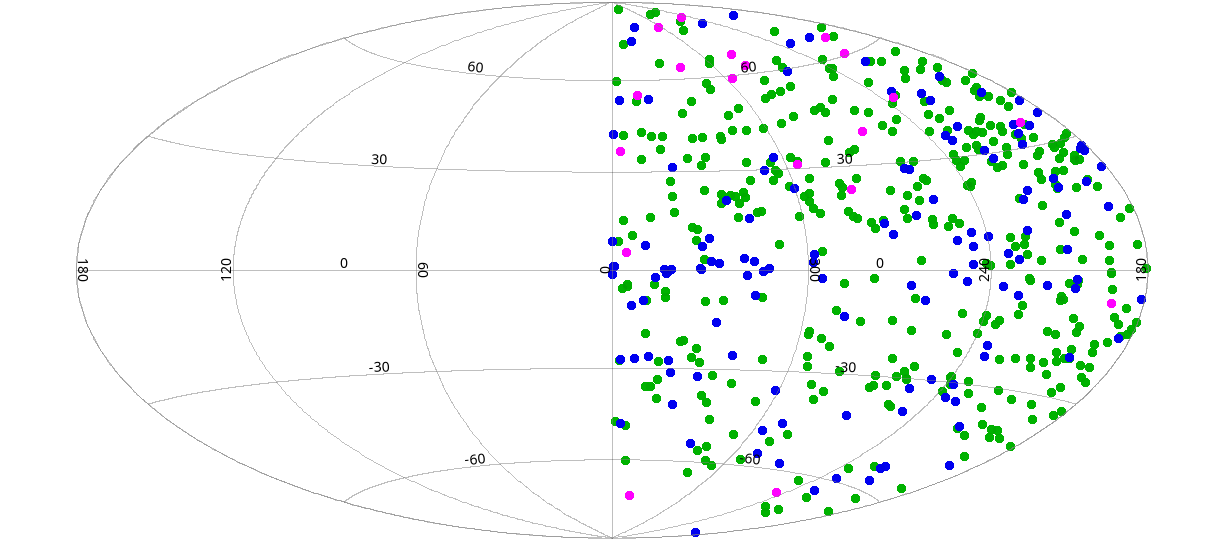}
 \includegraphics[width=0.54\columnwidth,trim=0.1cm 0cm 0.3cm 0.1cm,clip]{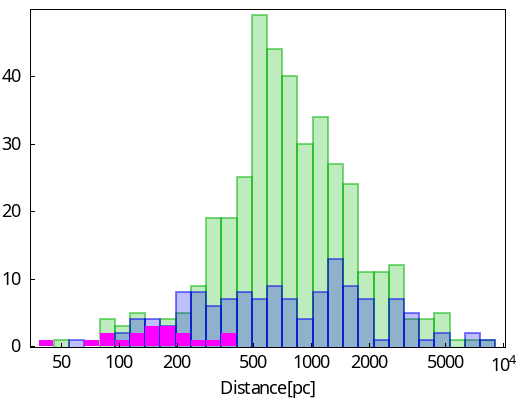}
  \caption {$537$ known CVs with an eROSITA counterpart, including $401$ non-magnetic CVs (in green) and $136$ magnetic CVs (in blue). $20$ of the known CVs have been classified as period-bouncers (in pink). We shown their distribution in the sky ({\it left panel}) and their distances ({\it right panel}).} \label{fig:eROSITACVdistribution}
 \end{center}
 \end{figure}

A cross-match between catalogs of known CVs (Ritter and Kolb 2003 \cite{ritter2003}; Drake et al. 2014 \cite{drake2014}; Pala et al. 2020 \cite{pala2020}; Inight et al. 2023a \cite{inight2023a}; Muñoz-Giraldo, Stelzer and Schwope 2024a \cite{munoz2024a}; Schwope 2025 \cite{schwope2025}; Catalog of Intermediate Polas and Candidates\footnote{https://asd.gsfc.nasa.gov/Koji.Mukai/iphome/catalog/alpha.html}) and the eRASS:4 catalog has lead to the identification of an X-ray counterpart for $537$ known CVs (see Fig.~\ref{fig:eROSITACVdistribution}). Of these X-ray detected CVs, $401$ are non-magnetic and $136$ are magnetic, implying that magnetic CVs represent $\sim$25$\%$ of the eROSITA-detected CV population. Amongst these known CVs we highlight the eROSITA detection of $20$ period-bouncers ($12$ of them classified as such based on their X-ray properties revealed by eROSITA; Muñoz-Giraldo, Stelzer and Schwope 2024a \cite{munoz2024a}; Muñoz-Giraldo, Stelzer and Schwope 2024b \cite{munoz2024b}). Therefore, period-bouncers represent at least $\sim$4$\%$ of the CV population detected by eROSITA.  

As discussed in Sect.~\ref{subsect:GRXE}, one of the central goals of eROSITA is to construct volume-limited samples of CVs, that are complete within 500pc for standard CVs and within 250pc for period-bounce CVs. Within a distance limit of 250pc, period-bouncers constitute at least $\sim$21$\%$ of the CV population detected by eROSITA. This result is in agreement with other observational surveys of CVs (Pala et al. 2020 \cite{pala2020}; Rodriguez et al. 2025 \cite{rodriguez2025}), serving as a starting point for the search of new period-bounce CVs with eROSITA.

\begin{figure}[h!]
\begin{center}
 \includegraphics[width=0.49\columnwidth,trim=0.1cm 0.1cm 0.4cm 0.3cm,clip]{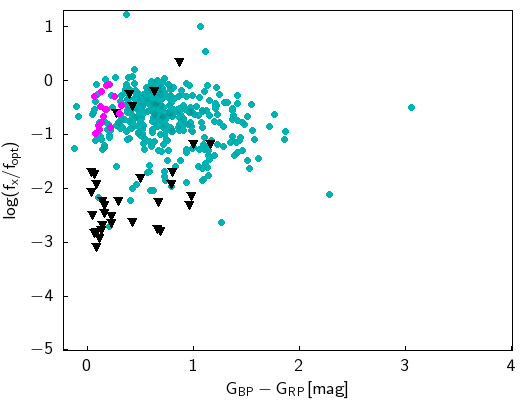}
 \includegraphics[width=0.49\columnwidth,trim=0.2cm 0.1cm 0.3cm 0.3cm,clip]{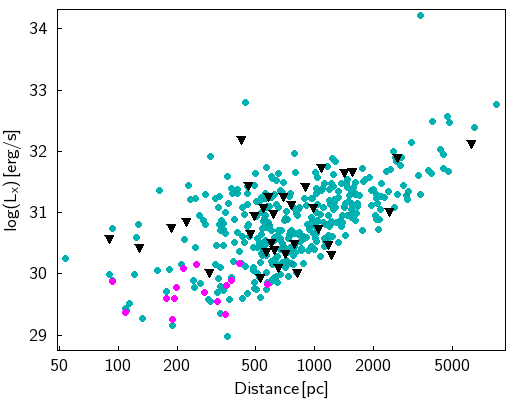}
  \caption {Diagnostic diagrams for the $401$ non-magnetic CVs with an eROSITA counterpart. We distinguish between nova-like variables (black triangles), dwarf novae (blue circles) and period-bouncers (pink circles). {\it Left panel: Gaia} color versus X-ray-to-optical flux ratio can be used to distinguish between nova-like variables and other types of CVs as the former are characterized by having higher mass accretion rates. {\it Right panel:} Distance versus X-ray luminosity. The lower envelope visible in this panel reflects the sensitivity limit of the eRASS:4 survey.} \label{fig:diagnosticNonM}
 \end{center}
 \end{figure}

\begin{figure}[h!]
\begin{center}
 \includegraphics[width=0.49\columnwidth,trim=0.1cm 0.1cm 0.4cm 0.3cm,clip]{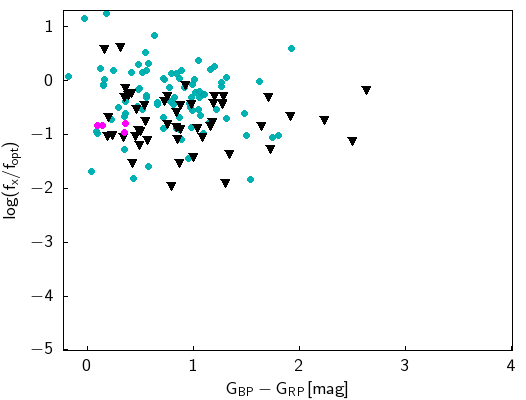}
 \includegraphics[width=0.49\columnwidth,trim=0.2cm 0.1cm 0.3cm 0.3cm,clip]{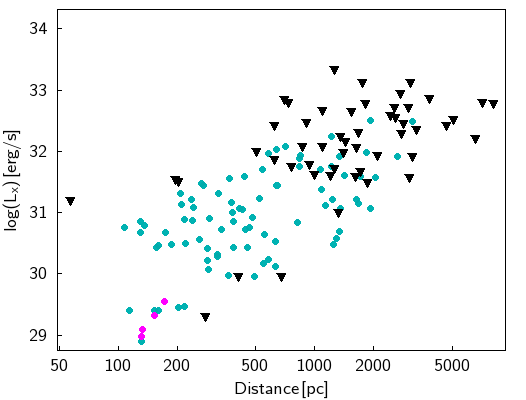}
  \caption {Diagnostic diagrams for the $136$ magnetic CVs with an eROSITA counterpart. We distinguish between intermediate polars (black triangles), polars (blue circles) and period-bouncers (pink circles). {\it Left panel: Gaia} color versus X-ray-to-optical flux ratio. {\it Right panel:} Distance versus X-ray luminosity can be used to distinguish between intermediate polars and other types of CVs as the former are characterised by having high X-ray luminosities. The lower envelope visible in this panel reflects the sensitivity limit of the eRASS:4 survey.} \label{fig:diagnosticM}
 \end{center}
 \end{figure}

A powerful tool for the identification of new CVs, and especially for differentiating between types of CVs, is the use of X-ray diagnostic diagrams (see Fig.~\ref{fig:diagnosticNonM} and Fig.~\ref{fig:diagnosticM}). These diagnostic diagrams can also help separate CVs from other kind of X-ray sources (X-ray binaries, AGN, coronal emitters among others). Non-magnetic CVs can be classified into the high mass accretion rate nova-like variables (typically found above the period gap) which exhibit very low X-ray-to-optical flux ratios, and the low mass accretion rate dwarf novae (typically found below the period gap) which exhibit higher flux ratios. Magnetic CVs are classified into polars, with strong magnetic fields ($\vert B \vert \gtrsim$ 10MG), exhibiting softer spectra and lower luminosities, and intermediate polars, with $|B|\lesssim10$ MG, showing higher X-ray luminosities.

Period-bouncers occupy a distinct region of parameter space, as they are characterized for being nearby, showing blue {\it Gaia} colors indicative of optical emission dominated by the WD, exhibiting low X-ray luminosities, and having X-ray-to-optical flux ratios consistent with extremely low mass accretion rates. Their distinct X-ray and optical properties make them excellent targets for systematic searches within the eROSITA data, particularly to search for the missing population of period-bouncers.

\subsection{Search for new systems}

The search for new CVs within the eROSITA all-sky survey is already on its way, which will lead to the construction of complete samples required to address the goals outlined in Sect.~\ref{subsect:goals}. Producing complete volume-limited samples, within 500pc for "standard" CVs and 250pc for the fainter period-bounce CVs, will allow detailed population studies. The identification of new CVs relies on combining their X-ray properties with optical and infrared information from large photometric and astrometric surveys, as well as on systematic follow-up campaigns to confirm CV candidates and determine their types.

\begin{figure}[h!]
\begin{center}
 \includegraphics[width=0.49\columnwidth,trim=0.1cm 0.1cm 0cm 0cm,clip]{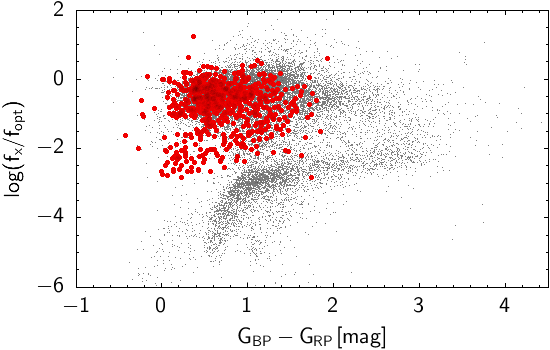}
 \includegraphics[width=0.50\columnwidth,trim=0.1cm 0.1cm 0cm 0cm,clip]{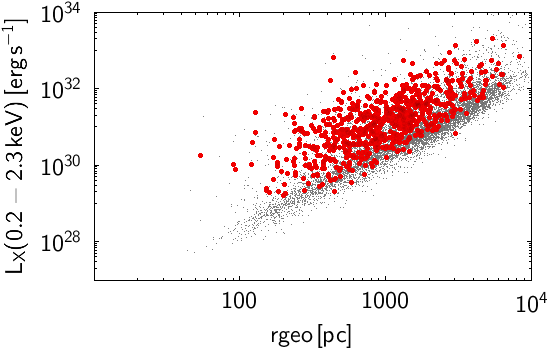}\\
 \includegraphics[width=0.50\columnwidth,trim=0.1cm 0.1cm 0cm 0.2cm,clip]{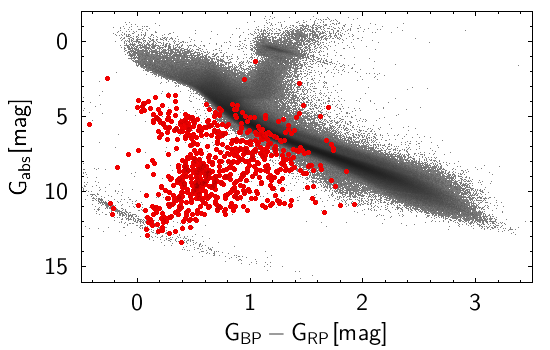}
 \includegraphics[width=0.49\columnwidth,trim=0.1cm 0.1cm 0cm 0cm,clip]{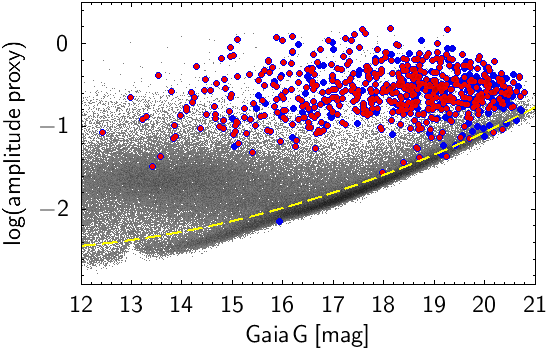}
  \caption {X-ray and optical characteristics of 11113 CV candidates from the eRASS:3 catalog (red) and other X-ray emitting sources (grey). {\it Top panels:} X-ray diagnostic diagrams. {\it Bottom left panel:} Hertzsprung–Russell diagram with CV candidates located mainly between the main-sequence and the WD locus. {\it Bottom right panel: Gaia} magnitude versus optical variability with CV candidates displaying larger optical variability than other X-ray emitting sources.} \label{fig:CVcandidates}
 \end{center}
 \end{figure}

\subsubsection{Cataclysmic variables}

An initial selection of CV candidates was performed using as a starting point the catalog compiling the information from the first three eROSITA all-sky surveys (eRASS:3). In order to identify the most likely optical counterparts, \texttt{NWAY}, a software for probabilistic cross-matching of catalogs (Salvato et al. 2018 \cite{salvato2018}), was used. The cross-matching combines positional information with a prior trained specifically on a representative sample of $624$ known CVs, incorporating their characteristic X-ray and optical properties (see Fig.~\ref{fig:CVcandidates}). This prior includes their distribution in the diagnostic diagrams introduced in Sect.~\ref{subsection:populations}, as well as additional optical diagnostics relating to brightness, colors, parallax, proper motion, sky position, and variability. This process ensures the presence of only a minor number of contaminant populations such as AGN or active late-type stars.

Applying \texttt{NWAY} with this CV-specific prior to the eRASS:3 catalog yielded 11113 sources with a probability of being a CV greater than 50$\%$. These candidates were prioritized for extensive spectroscopic follow-up using a wide range of facilities, including Sloan Digital Sky Survey (SDSS), Magellan, X-Shooter, the New Technology Telescope (NTT), and the Southern African Large Telescope (SALT), among others. This large-scale campaign aims to systematically verify the status of the candidate and, where applicable, classify the CV type.

\begin{figure}
\begin{center}
 \includegraphics[width=0.49\columnwidth,trim=0.1cm 0.1cm 0.2cm 0.2cm,clip]{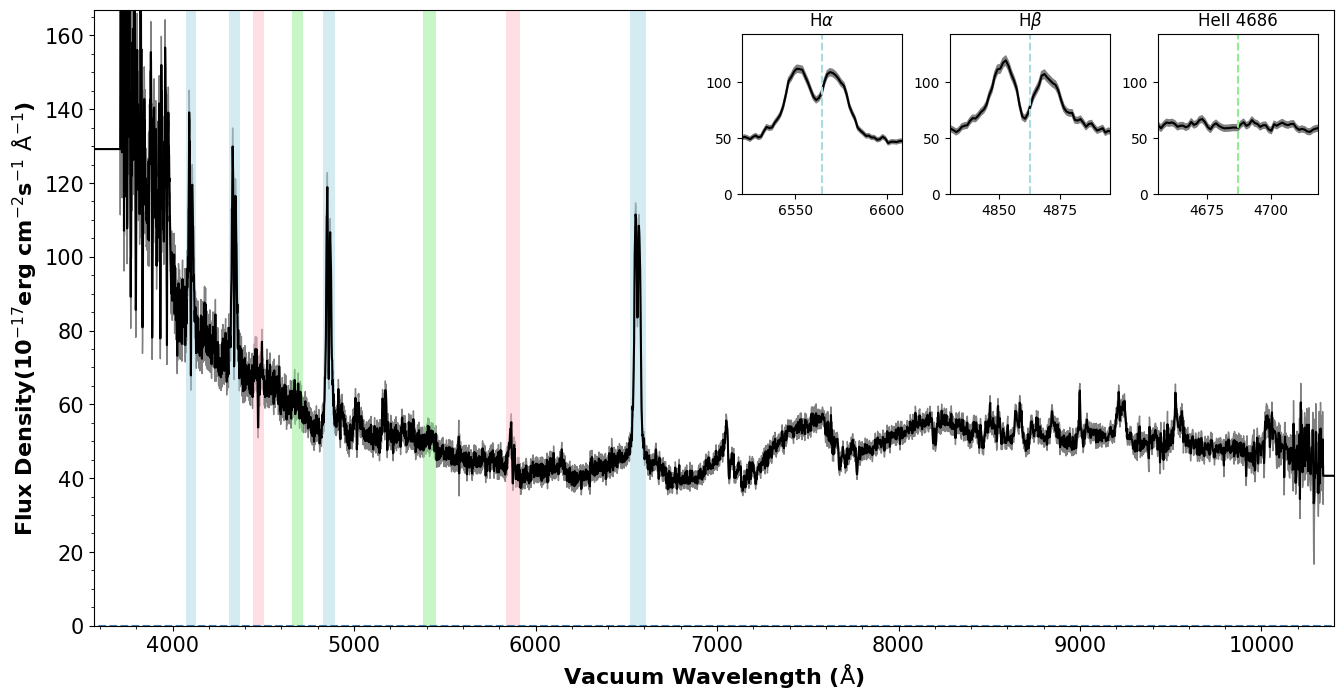}
 \includegraphics[width=0.49\columnwidth,trim=0.2cm 0.1cm 0.2cm 0.2cm,clip]{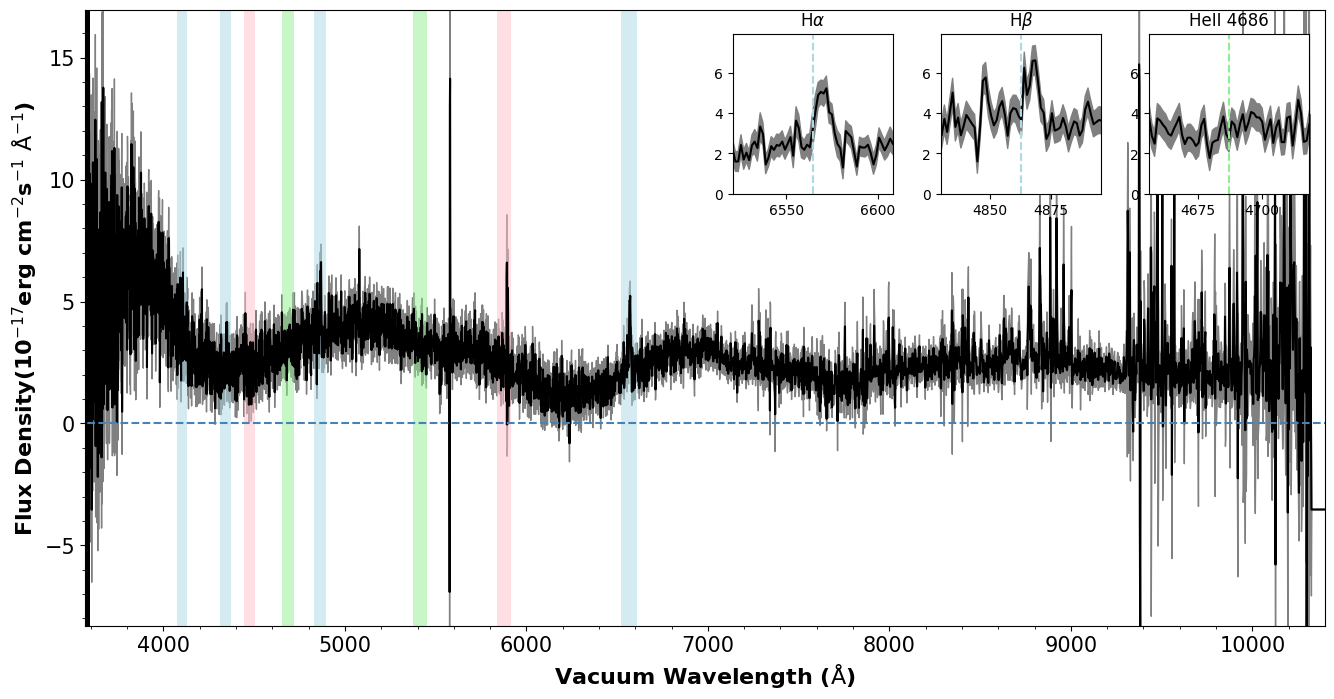}\\
 \includegraphics[width=0.49\columnwidth,trim=0.1cm 0.1cm 0.2cm 0.2cm,clip]{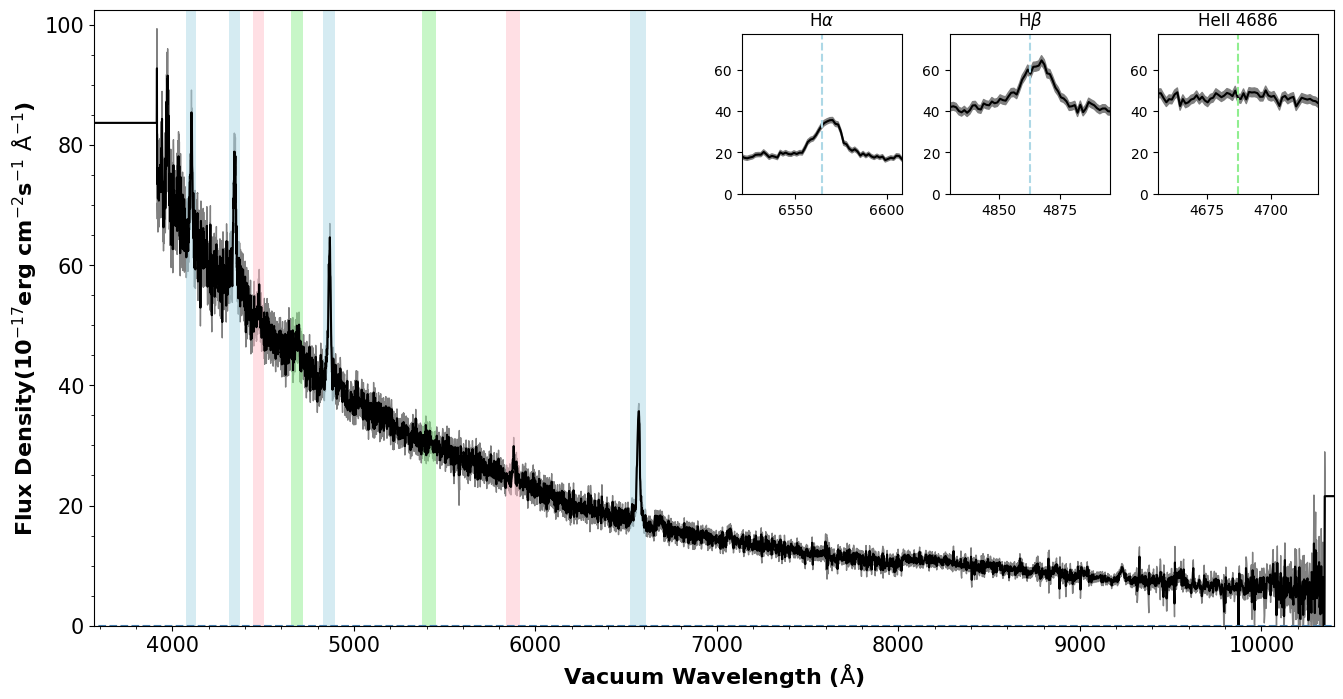}
 \includegraphics[width=0.49\columnwidth,trim=0.2cm 0.1cm 0.2cm 0.2cm,clip]{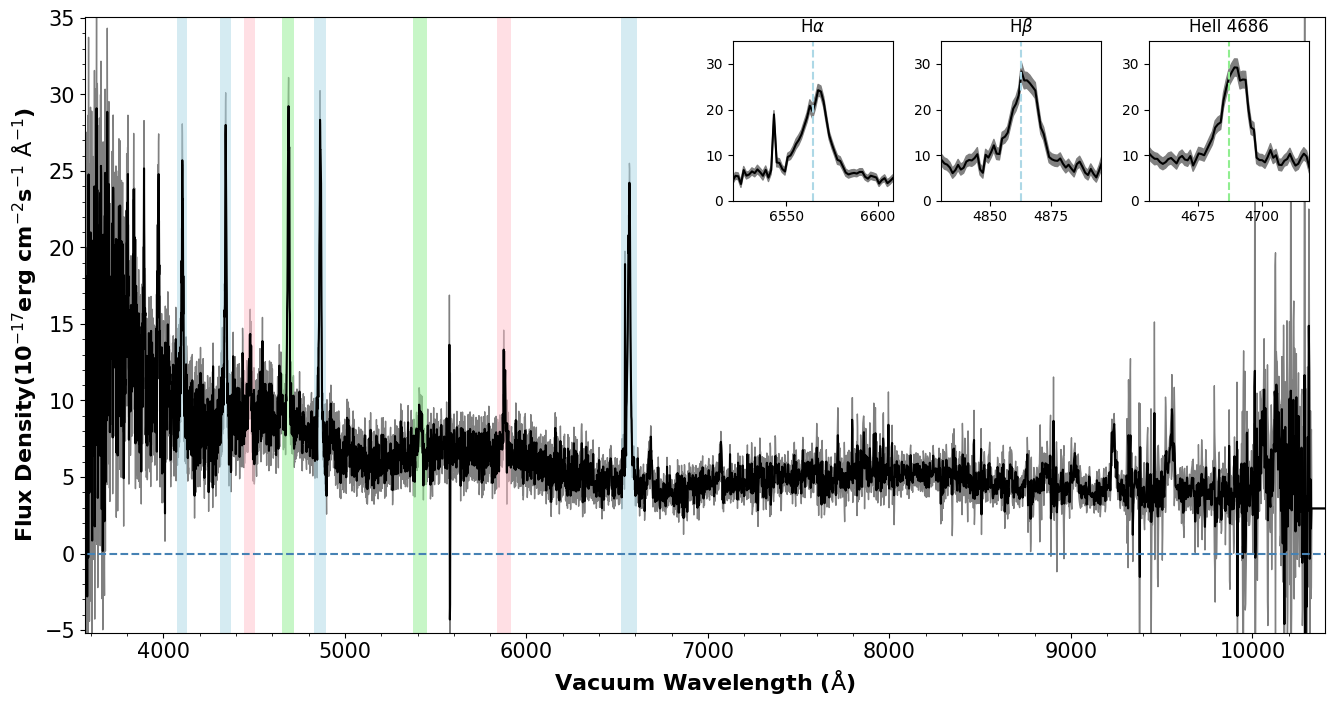}
  \caption {Optical spectra from the SDSS follow-up campaign of CV candidates. All these candidates were confirmed as CVs evidenced by the presence of Balmer emission lines (highlighted in blue). {\it Top left panel:} Low mass transfer CV with an M-type donor. {\it Top right panel:} Magnetic low mass transfer CV with a cyclotron continuum and weak lines. {\it Bottom left panel:} High mass transfer CV dominated by emission from a large disk. {\it Bottom right panel:} Magnetic high mass transfer CV with a cyclotron continuum.} \label{fig:sdssFollowup}
 \end{center}
 \end{figure}

A fraction of these candidates have already been confirmed as CVs. The confirmation relied primarily on the detection of prominent Balmer emission lines in their optical spectra (Fig.~\ref{fig:sdssFollowup}), indicative of ongoing accretion and therefore of a compact interacting binary (Warner 2003 \cite{warner2003}). The combination of follow-up spectroscopy and the X-ray information from eROSITA has also enabled a reliable classification of some of the confirmed systems into different CV types. This includes intermediate polars (identified through the detection of the WD spin period), low mass accretion rate CVs both non-magnetic (with spectral signatures of M dwarf donors) and magnetic (distinguished by cyclotron continuum features), and high mass accretion rate CVs in which the optical emission is dominated either by a bright accretion disk (non-magnetic) or by cyclotron harmonics (magnetic).

The successes and limitations identified during the eRASS:3 search phase helped improve the CV prior for the next selection of CV candidates based on eRASS:5. This improved matching process identified 30666 high-probability CV candidates. These sources will be the focus of spectroscopic and photometric follow-up campaigns, most notably through the upcoming 4MOST survey, which will provide uniform, high-quality spectroscopic coverage across a substantial fraction of the southern sky.

\subsubsection{Period-bounce cataclysmic variables}

Period-bouncers represent the most evolved CVs, reaching a point where the donor has become degenerate and the system exhibits a very low mass accretion rate. As a result, period-bouncers appear very similar to single WDs in the optical band, as the contribution from the low level accretion and degenerate donor are small compared to that of the WD.

Recent studies of known period-bouncers (Inight et al. 2023b \cite{inight2023b}; Muñoz-Giraldo et al. 2025 \cite{munoz2025}) have confirmed that these systems are indeed found in large WD catalogs (Kepler et al. 2016 \cite{kepler2016}; Jiménez-Esteban et al. 2018 \cite{jimenez2018}; Gentile Fusillo et al. 2019 \cite{gentile2019}; Gentile Fusillo et al. 2021 \cite{gentile2021}; Kepler et al. 2021 \cite{kepler2021}), suggesting that there might be more period-bouncers classified as single WDs and hiding in WD catalogs.

A key observational difference between genuine single WDs and period-bouncers is the presence of an X-ray emission. Period-bouncers exhibit detectable X-ray emission produced by low level mass accretion onto the WD surface, while single WDs are only expected to emit detectable X-rays if they are significantly hotter (T${\rm eff}\approx25,000$ K; Friedrich et al. 2025 \cite{friedrich2025}) than the cool temperatures typical of evolved CVs (T${\rm eff}\lesssim12,500$ K; Pala et al. 2022 \cite{pala2022}). Therefore, the combination of a cool WD–like optical appearance with a detected X-ray counterpart is a strong indicator that an object may be a period-bounce CV.

Identifying period-bouncers, however, remains challenging. These systems do not occupy a unique locus in standard diagnostic diagrams and often overlap with the parameter space of single WDs or low mass accretion rate CVs. Consequently, we rely on a multi-wavelength period-bouncer scorecard (Muñoz-Giraldo, Stelzer and Schwope 2024a \cite{munoz2024a}), which assigns each candidate a score based on its consistency with known period-bouncer characteristics. The scorecard includes ten parameters:

\begin{enumerate}
    \item {\it Donor spectral type:} The most important parameter of the scorecard, since a direct spectroscopic detection of an L or T type donor provides the strongest confirmation of a period-bouncer. Photometric signatures consistent with very late-type donors are also considered.
    \item {\it Donor mass:} A mass below 0.058 M$_\odot$ indicates a degenerate donor (Knigge, Baraffe, and Patterson 2011 \cite{knigge2011}).
    \item {\it Orbital period:} This parameter is evaluated together with donor mass, as the orbital period alone is not a reliable indicator of evolutionary status near the period minimum.
    \item {\it WD temperature:} Period-bouncers, that are not in outburst, host cool WDs due to sustained low mass accretion rates.
    \item {\it Gaia variability:} Minimal optical variability is expected because outbursts or superoutbursts have recurring times for period-bouncers on $\sim$10000 day timescales (Patterson 2011 \cite{patterson2011}).
    \item {\it Gaia colors:} Location consistent with the WD locus due to the WD-dominated optical emission.
    \item {\it SDSS colors:} Similar to {\it Gaia colors} the location of period-bouncers is expected to overlap with the WD sequence.
    \item {\it Ultraviolet colors:} As GALEX colors are reflective of the WD temperature, they should indicate a cool WD in quiescence.
    \item {\it Infrared colors:} Location consistent with observed WD/brown dwarf binaries.
    \item {\it Infrared excess:} In systems with very late-type donors, the spectral energy distribution reveals a small excess over the WD emission, making this a key discriminator relative to single WDs which do not present any excess.
\end{enumerate}

These scorecard criteria are supplemented with X-ray selection cuts designed to ensure that candidates exhibit X-ray properties consistent with known period-bouncers. Specifically, we require an X-ray-to-optical flux ratio between -0.9 and 0.3, characteristic of low mass accretion rate systems, and an X-ray luminosity below 3.3$\times$10$^{30}$erg/s, expected for faint and evolved CVs. Together, the scorecard and X-ray selection cuts aim to ensure that each candidate is consistent with a system dominated by a cool WD in the optical and ultraviolet band, with evidence of a late-type donor from the infrared emission, and an X-ray emission consistent with faint, low mass accretion rates systems. Literature values for the spectral type of the donor, the donor mass and the orbital period of the candidates are used for the confirmation when available.

To date, these tools have already been used to confirm $12$ new period-bouncers, mainly thanks to the eROSITA detection of low mass accretion rates consistent with highly evolved CVs (Muñoz-Giraldo, Stelzer and Schwope 2024a \cite{munoz2024a}; Muñoz-Giraldo, Stelzer and Schwope 2024b \cite{munoz2024b}). This almost doubles the previously known population of period-bouncers and highlights the potential of eROSITA for expanding the population by identifying new candidates.

We applied a reduced version of our period-bouncer scorecard to the $601$ WDs from the catalog by Gentile Fusillo et al. (2021) \cite{gentile2021} that have an eROSITA counterpart and are located within 500pc, in order to identify the most likely period-bouncer candidates. Reducing the scorecard was necessary because most of these sources lack measurements of key binary parameters such as orbital period, donor mass, or donor spectral type. As shown by Muñoz-Giraldo et al. 2025 \cite{munoz2025}, the most useful and relevant parameters in our scorecard for the identification of period-bouncers are the donor spectral type, {\it Gaia} colors, and ultraviolet colors. Since the reduced version incorporates two of these three key parameters, it remains a robust identification tool. 

\begin{figure}
\begin{center}
 \includegraphics[width=0.49\columnwidth,trim=0.1cm 0.1cm 0cm 0.2cm,clip]{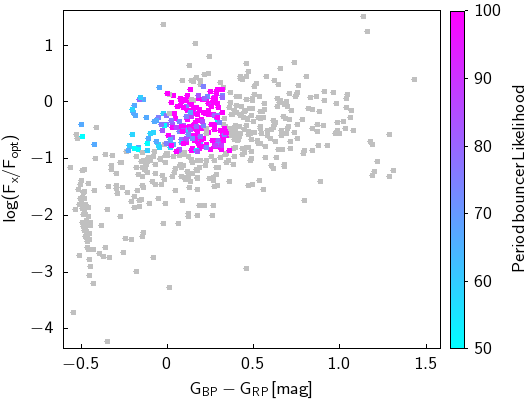}
 \includegraphics[width=0.49\columnwidth,trim=0.2cm 0.1cm 0cm 0.2cm,clip]{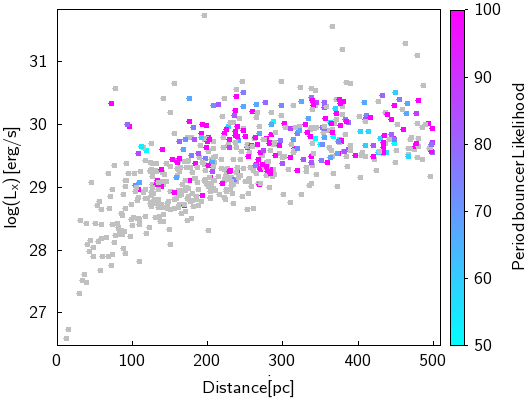}
  \caption {Distribution of the $601$ WDs from Gentile Fusillo et al. (2021) \cite{gentile2021} with an eROSITA counterpart in the X-ray diagnostic diagrams used to identify period-bounce CVs. The final score for the systems that fulfill both X-ray selection cuts is shown as a color scale.} \label{fig:GFscores}
 \end{center}
 \end{figure}

Out of the $601$ WDs with an eROSITA counterpart, $213$ fulfill the X-ray selection cuts and therefore qualify as period-bouncer candidates (shown with a color scale in Fig.~\ref{fig:GFscores}). Systems that do not fulfill the X-ray selection cuts are shown in grey and excluded from further consideration.

When the reduced scorecard is applied to all known eROSITA-detected period-bouncers, the lowest score obtained is 72$\%$ (for V406 Vir). We adopt this value as the threshold for identifying high-likelihood period-bouncer candidates. Using this criteria, $161$ of the $213$ sources are classified as high-likelihood period-bounce candidates.

A cross-match with SIMBAD\footnote{SIMBAD stands for Set of Identifications, Measurements and Bibliography for Astronomical Data, available at \url{http://simbad.cds.unistra.fr/simbad.}} as well as an extensive literature search revealed that out of these $161$ high-likelihood period-bounce candidates: $16$ were already known period-bouncers (including $9$ recently confirmed with eROSITA; Muñoz-Giraldo, Stelzer and Schwope 2024a \cite{munoz2024a}; Muñoz-Giraldo, Stelzer and Schwope 2024b \cite{munoz2024b}), $5$ are classified as CVs, $2$ are known AM CVn binaries, $1$ is a RR Lyrae variable candidate, and $137$ are previously known WDs or WD candidates. These $137$ high-likelihood period-bounce candidates that are known in the literature as WDs or WD candidates represent a substantial reservoir for the identification of new period-bouncers.

The follow-up campaigns required to confirm these candidates are already on their way, with instruments like SDSS, X-Shooter, among others. Among the candidates selected for follow-up, two systems stand out as particularly intriguing examples of evolved CVs: a newly confirmed period-bouncer (GALEX J125751.4–283015; Muñoz-Giraldo et al. 2025 \cite{munoz2025}), and a potential detached period-bouncer. 

\begin{figure}
    \includegraphics[width=0.54\textwidth]{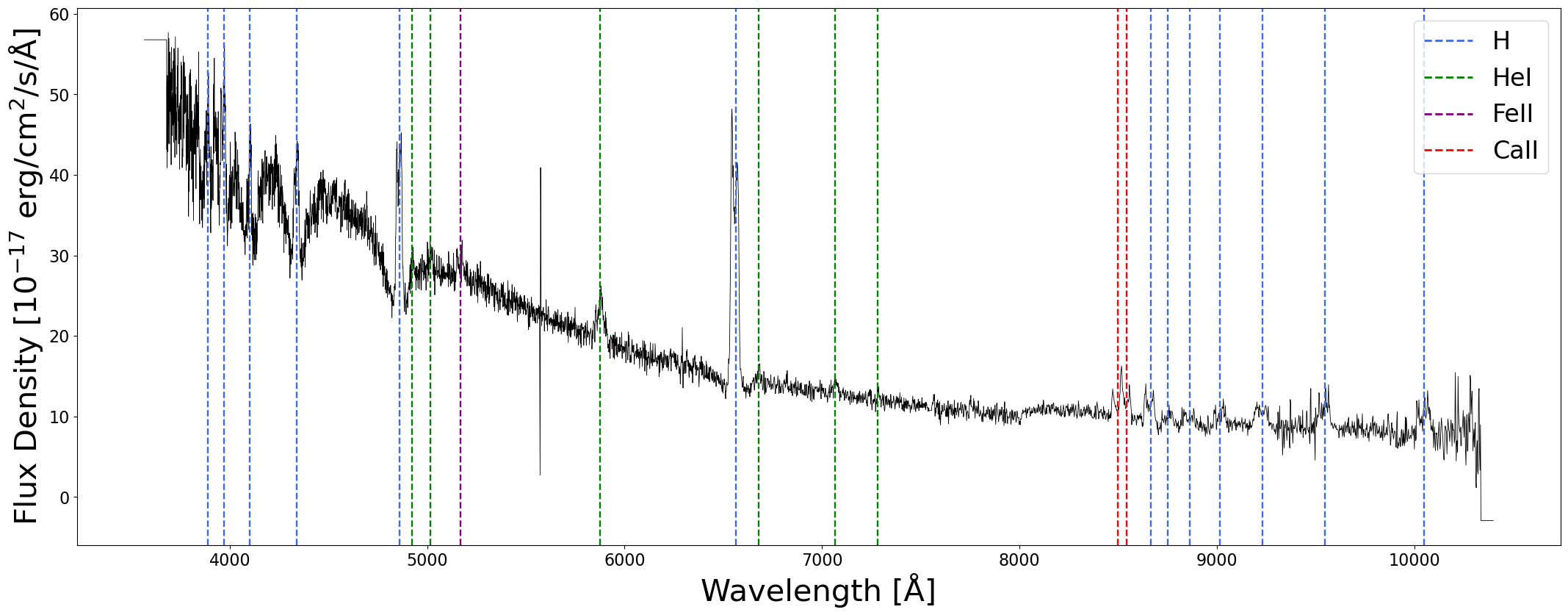}\hfill
    \includegraphics[width=0.45\textwidth,trim=1.7cm 0.2cm 3.7cm 2.1cm,clip]{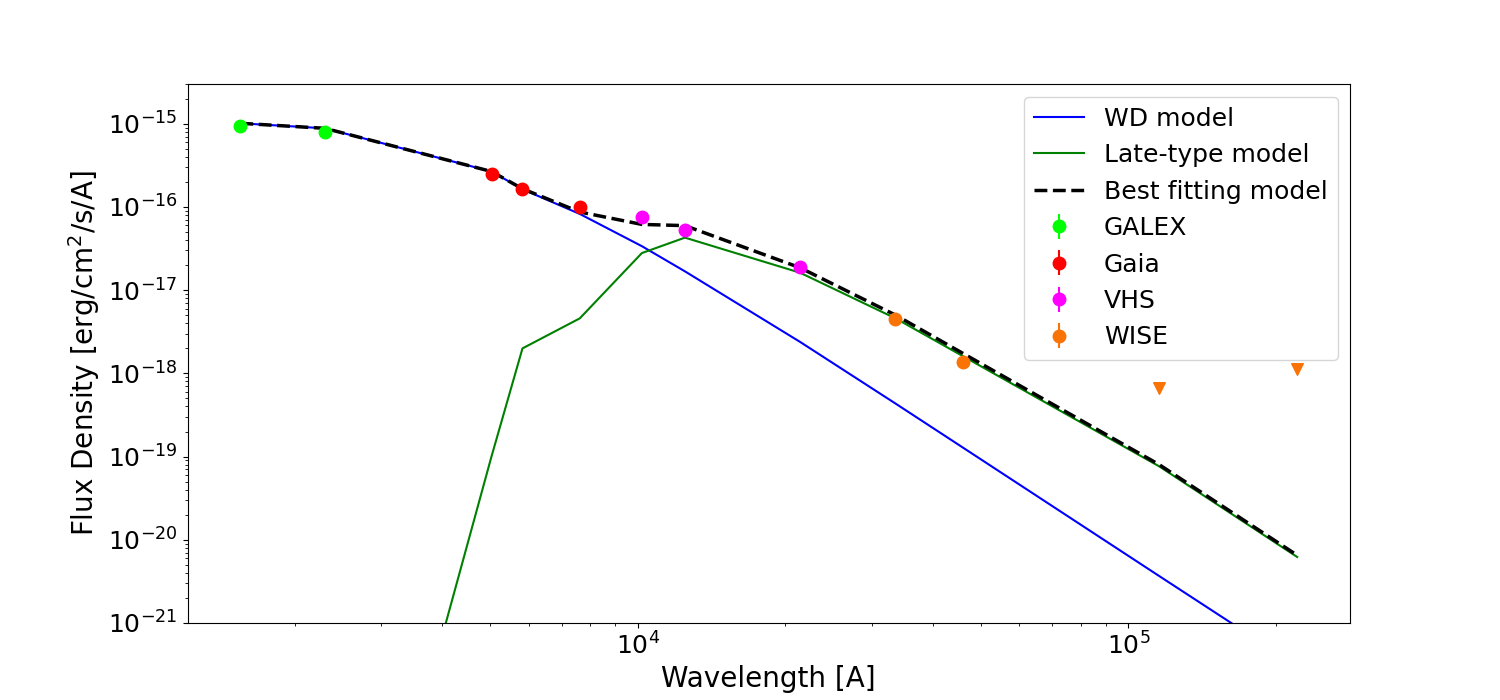}
    \includegraphics[width=\textwidth,trim=6.2cm 0cm 5.6cm 1.4cm,clip]{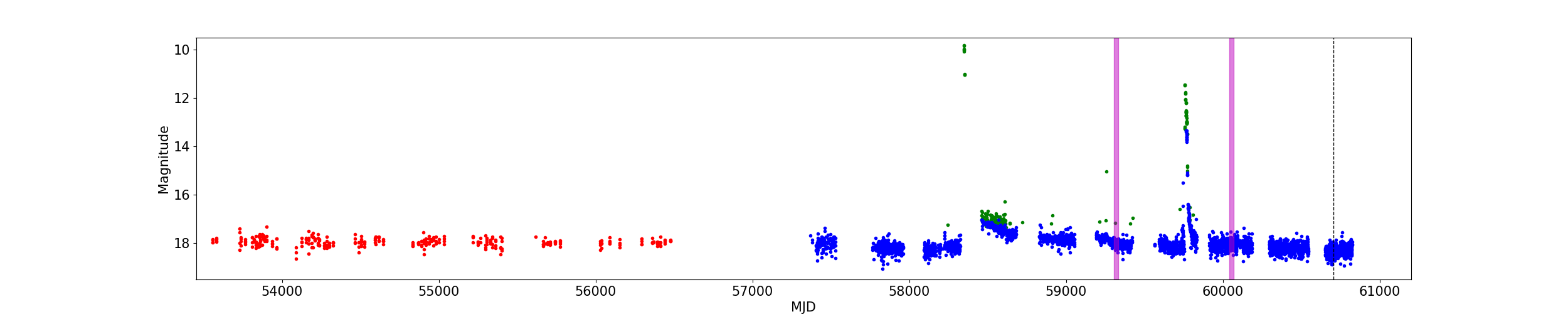}
    \caption{Newly confirmed period-bouncer GALEX J125751.4-283015. \textit{Upper left panel:} SDSS spectrum showing clear Balmer emission lines. \textit{Upper right panel:} Spectral energy distribution fitted using a hydrogen WD model (T$_{\rm eff}$ = 12000\,K and $\log(g)$=8.0) and a late-type donor model (T$_{\rm eff}$ = 2000\,K and $\log(g)$=4.5). \textit{Lower panel:} Combined light curves from CRTS (red), ATLAS (blue) and ASAS-SN (green). Vertical purple stripes show the epochs of the TESS light curves, while the vertical dashed black line shows the the epoch of the SDSS spectrum. Adapted from Muñoz-Giraldo et al. 2025 \cite{munoz2025}.}
    \label{fig:GALEX1257}
\end{figure}

GALEX J125751.4–283015 is listed in SIMBAD as a WD candidate and it has been previously associated with the transient ASASSN-18su (Barba 2018 \cite{barba2018}; Le Guillou et al. 2015 \cite{leGuillou2015}). The SDSS-V spectrum shown in Fig.~\ref{fig:GALEX1257} confirms its CV nature exhibiting Balmer emission lines as well as the accompanying Paschen, HeI, CaII, and FeII lines characteristic of quiescent dwarf novae (Mason and Howell 2005 \cite{mason2005}; Southworth et al. 2009 \cite{southworth2009}). Combining the SDSS-V spectrum together with GALEX photometry yields a WD temperature between 12000K and 12500K and a $\log g=8.0$–$8.5$, consistent with CVs near the period minimum (Pala et al. 2020 \cite{pala2020}; Pala et al. 2022 \cite{pala2022}), and that imply a WD mass of $0.57$\,$M_\odot$ to 0.67\,$M_\odot$ (Nauenberg 1972 \cite{nauenberg1972}).

Short-cadence TESS data reveals significant periodicities at 41.26\,min and 82.17\,min, both above the detection threshold of Hernández-Díaz et al. (2025) \cite{hernandez2025}. We interpret the latter as the orbital period and the former as its harmonic. Such double-humped modulations are common in short-period CVs and may arise from ellipsoidal variations, partial eclipses, or disk structures associated with the 2:1 resonance in low-mass ratio systems (Bochkarev, Karitskaia and Shakura 1979 \cite{bochkarev1979}; Froning et al. 1999 \cite{froning1999}; Skidmore et al. 2000 \cite{skidmore2000}; Aviles et al. 2010 \cite{aviles2010}; Zharikov et al. 2013 \cite{zharikov2013}; Amantayeva et al. 2021 \cite{amantayeva2021}).

Archival photometry from CRTS, ATLAS, and ASAS-SN (see Fig.~\ref{fig:GALEX1257}) shows that the system has remained at $\sim$18mag for most of the past two decades, with only two recorded outbursts. Such infrequent activity is typical of low mass accretion rate systems like period-bouncers. Because both the TESS and SDSS data were obtained during quiescence, they reliably reflect the binary parameters of the system.

A multi-wavelength SED built from GALEX, Gaia, VHS, and AllWISE photometry (following Muñoz-Giraldo et al. 2024a \cite{munoz2024a}) confirms a WD with $T_{\rm eff}=12000\pm125$K and $\log g=8.0\pm0.125$, and reveals an IR excess consistent with an evolved donor. A binary fit using BT-Settl models (Allard et al. 2003 \cite{allard2003}) yields a secondary with $T_{\rm eff}=2000\pm25$K and $\log g=4.5\pm0.25$ (Fig.~\ref{fig:GALEX1257}), corresponding to a spectral type of L0 (Knigge et al. 2011 \cite{knigge2011}), but because we expect some contribution from the accretion disc we characterize the donor as having a spectral type of L0 or later (Pala et al. 2018 \cite{pala2018}; Neustroev and Mäntynen 2023 \cite{neustroev2023}).

Overall, GALEX J125751.4–283015 is an evolved, short-period CV with a low X-ray luminosity and a late-type donor, firmly identifying it as a new period-bouncer. Future infrared spectroscopy will enable direct detection of the donor and confirm its status as a degenerate star.

\begin{figure}
    \centering
    \includegraphics[width=0.47\textwidth,trim=0.2cm 0.1cm 0cm 1.2cm,clip]{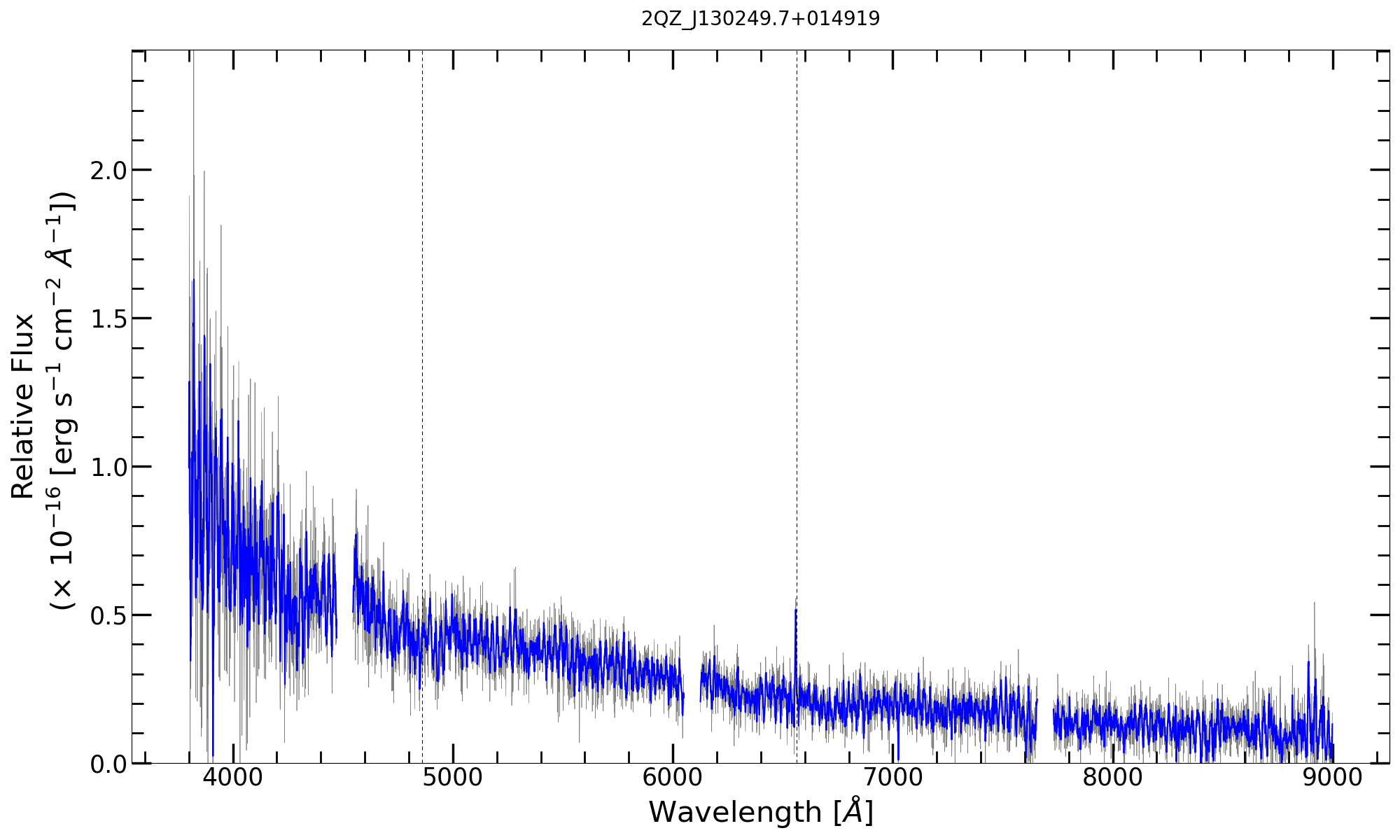}\hfill
    \includegraphics[width=0.52\textwidth,trim=1.8cm 0.2cm 2.4cm 1.4cm,clip]{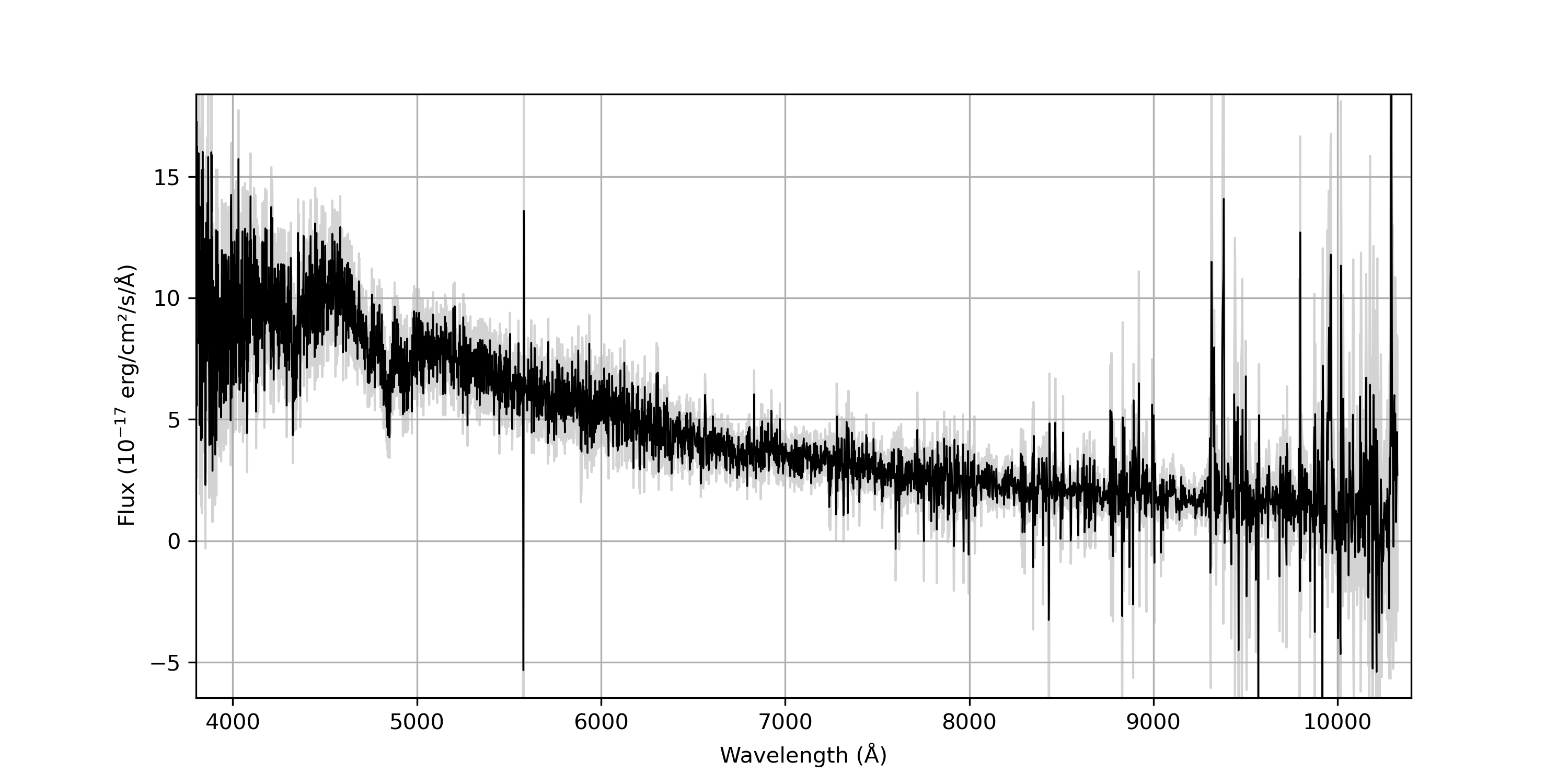}
    \includegraphics[width=0.55\textwidth,trim=1.7cm 0.2cm 3.7cm 2.1cm,clip]{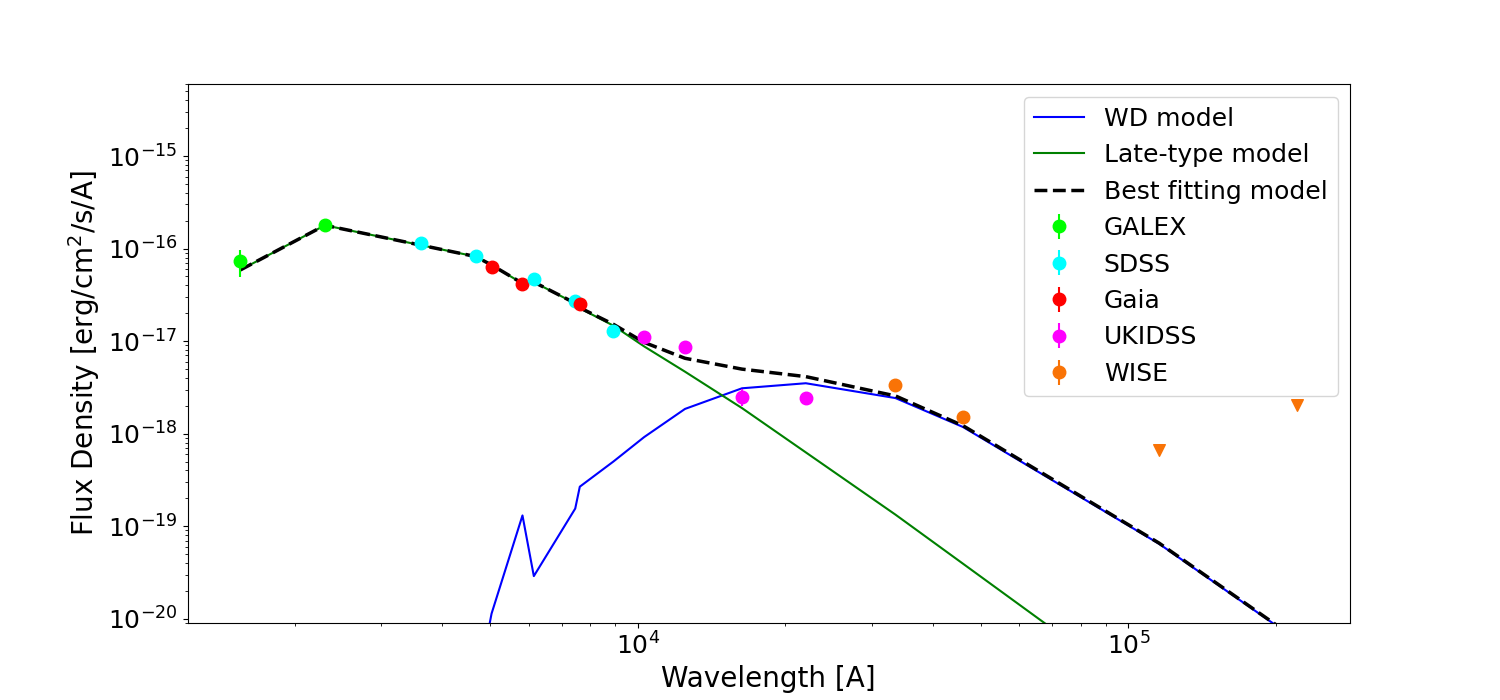}
    \caption{Detached period-bouncer candidate. \textit{Upper left panel:} Magellan spectrum showing a weak H$\alpha$ emission line. \textit{Upper right panel:} SDSS spectrum with an absent H$\alpha$ emission line. \textit{Lower panel:} Spectral energy distribution fitted using a hydrogen WD model (T$_{\rm eff}$ = 10750\,K and $\log(g)$=8.5) and a late-type donor model (T$_{\rm eff}$ = 1500\,K and $\log(g)$=3.5).}
    \label{fig:2QZJ1302}
\end{figure}

The other candidate appears to be a promising detached period-bouncer, with two independent spectra that reveal a H$\alpha$ emission line that is either very weak (Fig.~\ref{fig:2QZJ1302} upper left panel) or almost absent (Fig.~\ref{fig:2QZJ1302} upper right panel). This is indicative of an extremely low mass accretion rate or possibly absent mass transfer that would be consistent with the detached period-bouncers proposed by Schreiber, M. R. and Belloni, D. 2025 \cite{schreiber2025}. Additionally, the spectral energy distribution of this system suggests the presence of a very late-type donor, making it a compelling detached period-bouncer candidate (Fig.~\ref{fig:2QZJ1302}). Follow-up with X-Shooter and XMM-Newton has already been submitted, which will help clarify the evolutionary status and assess whether this candidate represents the first ever observed detached period-bouncer.

\section{Summary and conclusions}

CVs represent a key class of interacting binaries, with an evolution that is driven by angular momentum loss and the progressive mass loss from the donor. As reviewed in Sect.~\ref{sect:CVevolution}, theoretical models predict a well-defined evolutionary sequence that goes from long-period, high mass accretion rate CVs to short-period CVs hosting cool WDs and, ultimately, degenerate donors at the period-bounce stage, and the hypothetical, as yet unobserved post-bounce detached state. While this overall picture is well established, major open questions remain, including the space density of magnetic and non-magnetic CVs, the role of WD magnetism in shaping binary evolution, the long-standing discrepancy between the predicted and observed fractions of period-bounce CVs, and the identification of post-bounce detached systems.

In Sect.~\ref{sect:xmm}, we highlighted the unique contribution of {\it XMM-Newton} to the study of CVs, particularly for magnetic systems. The high time resolution and sensitivity of {\it XMM-Newton} has enabled detailed characterizations of polars, including their accretion geometry, orbital parameters, and accretion regimes. The detection and analysis of period-bounce polars, with their intrinsically faint and soft X-ray spectra, further demonstrate the critical role of pointed X-ray observations in identifying and characterizing the most evolved CVs.

Sect.~\ref{sect:eROSITA} focused on the large impact of the eROSITA all-sky survey. eROSITA has already detected X-ray emission from more than 3 million sources, yielding the largest X-ray catalog to date, specially in the soft band. Cross-matching this catalog with existing CV compilations resulted in the identification of $537$ known CVs in the German eROSITA sky, including 20 period-bouncers, of which twelve were newly classified using eROSITA data. Diagnostic diagrams combining X-ray, optical, and astrometric information provide an efficient pathway for separating CVs from other X-ray emitting populations and for distinguishing between CV subtypes.

A major long-term objective of the eROSITA CV program is the construction of volume-limited samples that are complete up to 500pc for standard CVs and up to 250pc for period-bouncers. These samples will allow statistical studies of CV evolution, the magnetic and non-magnetic subpopulations, and the contribution of CVs to the GRXE. The first systematic search for new CVs using eRASS:3 and a Bayesian cross-matching approach has already yielded more than ten thousand high-probability candidates, several of which have been spectroscopically confirmed. For period-bouncers, a dedicated multi-wavelength scorecard applied together with X-ray selection cuts has proven highly effective, identifying more than a hundred promising candidates and doubling the number of known period-bouncers.

The early confirmation results illustrate the potential of the combining an eROSITA-selection of candidates with photometric and, specially, spectroscopic follow-up campaigns. A newly confirmed period-bouncer shows a cool WD, an L0 or later type donor, low mass accretion luminosity and an 82\,min orbital period, consistent with a highly evolved system. Another candidate may represent the first detached period-bounce CV uncovered by eROSITA, with spectroscopy indicating extremely weak or absent accretion signatures. Continued follow-up with facilities such as SDSS, X-Shooter, and {\it XMM-Newton} will be crucial to verify these classifications.

Overall, eROSITA, together with support from ongoing photometric, spectroscopic and ultraviolet surveys, opens a new window onto the CV population. Its sensitivity and sky coverage provide the means to tackle long standing questions in CV evolution, to uncover large numbers of previously unidentified systems, and to build statistically robust samples of both standard and period-bounce CVs. As the follow-up campaigns progress, these efforts will shape our understanding of compact binary evolution and the role of accreting WDs within the Galaxy.

\section{Acknowledgements}

DMG would like to thank Paula Kvist for fruitful discussions on observations of cataclysmic variables and period-bouncers during this workshop. DMG would like to also thank Gloria Sala, Isabel Lima, Nikita Rawat and Na'ama Hallakoun for valuable discussions about cataclysmic variables and binaries in a more general perspective. DMG was supported by Deutsche Forschungsgemeinschaft (DFG) under grant number STE 1068/6-1.

\bigskip
\bigskip
\noindent {\bf DISCUSSION}

\bigskip
\noindent {\bf SERGEY ZHARIKOV:} How do you define period-bouncer systems? In general, the presence of X-ray alone is not sufficient to make such a claim, nor is the detection of Balmer lines from a WD. X-rays typically indicate ongoing accretion, while the WD spectrum points to an optical thin accretion disk. A reliable classification as a period-bouncer requires an orbital period longer than about 85 minutes (given the uncertainties in determining a period minimum) together with estimates of the accretion rate and mass ratio between the components. 

\bigskip
\noindent {\bf DANIELA MUÑOZ GIRALDO:} Period-bouncers are defined as CVs that have evolved past the period minimum, containing a brown dwarf donor. I agree that in order to be thorough in the confirmation of systems as period-bouncers there should be an estimate of the mass ratio between components. However, we consider that the data/observations required to obtain this value (especially for non-eclipsing systems) is very difficult to obtain for large samples of candidates. That is why our approach centers on identifying new period-bouncers based on more readily available data (like photometry, X-ray emission from eROSITA or spectroscopy from SDSS) that allows to confirm the number of new period-bouncers that models expect.
\end{document}